\pdfoutput=1
\documentclass[journal]{IEEEtran}

\usepackage{cite}
\usepackage{amsmath,amssymb,amsfonts}
\usepackage{algorithm}
\usepackage{algorithmic}
\usepackage{graphicx}
\usepackage[percent]{overpic}
\usepackage{textcomp}
\usepackage{xcolor}
\usepackage[colorlinks,linkcolor=blue,citecolor=blue,urlcolor=blue]{hyperref}
\usepackage{bm}

\newcommand{\revb}[1]{#1}

\newtheorem{remark}{Remark}

\usepackage{titlesec}

\titlespacing{\subsection}{0pt}{1ex plus .2ex minus .2ex}{1ex plus .2ex}

\IEEEoverridecommandlockouts
\IEEEaftertitletext{\vspace{-1\baselineskip}}

\begin{document}

\title{A Unified Two-Stage Generative Diffusion Framework for Channel Estimation and Port Selection in Multiuser MIMO-FAS}

\author{
    Erqiang~Tang,
    Wei~Guo,~\IEEEmembership{Member,~IEEE,}
    Hengtao~He,~\IEEEmembership{Member,~IEEE,}
    Shenghui~Song,~\IEEEmembership{Senior Member,~IEEE,}
    Jun~Zhang,~\IEEEmembership{Fellow,~IEEE,}
    and~Khaled~B.~Letaief,~\IEEEmembership{Fellow,~IEEE}%
    \thanks{This work has been submitted to the IEEE for possible publication. Copyright may be transferred without notice, after which this version may no longer be accessible.}%
    \thanks{Erqiang Tang, Wei Guo, Shenghui Song, Jun Zhang, and Khaled B. Letaief are with the Department of Electronic and Computer Engineering, The Hong Kong University of Science and Technology (HKUST), Hong Kong. Emails: etangaa@connect.ust.hk, \{eeweiguo, eeshsong, eejzhang, eekhaled\}@ust.hk.}%
    \thanks{Hengtao He is with the National Mobile Communications Research Laboratory, Southeast University, Nanjing, China. Email: hehengtao@seu.edu.cn.}%
}

\maketitle

\begin{abstract}
Fluid antenna systems (FAS) have emerged as a promising technology for next-generation wireless systems. However, practical multiuser multiple-input multiple-output FAS (MIMO-FAS) faces two inherently coupled challenges: acquiring accurate high-dimensional channel state information (CSI) from limited RF chains and solving the combinatorial port selection problem, where the effectiveness of the latter highly depends on the result of the former. In this paper, we propose a unified two-stage diffusion framework that formulates the joint task as a maximum-a-posteriori (MAP) inference problem \revb{equivalent to maximizing the posterior expected utility of the port selection, and approximates it through two sequential stages}. For Stage~I, we employ a continuous flow-based diffusion model as a powerful implicit prior for 2D FAS channels and perform approximate posterior sampling via a parallel guided generation scheme, enabling accurate multiuser channel recovery even under severely low sub-sampling ratios. \revb{For Stage~II, we train a discrete diffusion model on optimal labels computed by exhaustive search to propose a pool of promising port selection candidates. The final selection is then determined by maximizing the empirical posterior expected utility over this pool, followed by a lightweight local refinement.} Extensive simulations demonstrate that the proposed framework simultaneously achieves \revb{accurate channel estimation and delivers over $90\%$ of the deployed rate of the computationally expensive exhaustive search over the estimated channel, while substantially improving the minimum achievable rate relative to practical cascaded schemes.}
\end{abstract}


\section{Introduction}

\IEEEPARstart{D}{riven} by the proliferation of edge devices and wireless applications, the forthcoming sixth-generation (6G) wireless networks are anticipated to offer tremendous data rates, ultra-low latency, and robust reliability~\cite{letaief2021edge,saad2019vision}. To satisfy these stringent requirements, massive multiple-input multiple-output~(MIMO) has been widely regarded as a key enabling technology~\cite{wang2023road}. However, as conventional MIMO architectures scale up to accommodate the next-generation demands, they encounter two fundamental challenges. First, equipping extremely large antenna arrays with a massive number of radio frequency (RF) chains imposes prohibitive hardware costs and significant power consumption~\cite{wang2024tutorial}. Second, conventional MIMO systems rely on fixed-position antenna~(FPA) arrays that are constrained by the half-wavelength antenna spacing rule. This fixed arrangement restricts their capability to fully exploit the spatial variations inherent in wireless channels, thereby limiting the overall system performance in terms of diversity and/or multiplexing gains~\cite{shojaeifard2022mimo}.

To address these limitations, the fluid antenna system~(FAS) has recently been proposed~\cite{wong2020fluid, wong2020fluid2}. Conceptually, FAS encompasses any form of movable or non-movable antennas that are capable of reconfiguring their positions. The possibility of adjusting antenna positions (also referred to as `ports') provides additional degrees of freedom (DoF) in space, thus allowing FAS to better explore the spatial diversity within a predefined region.

While initial investigations on FAS mainly focused on single-input single-output~(SISO) setups where only a single port is activated~\cite{wong2020performance, new2023fluid, tlebaldiyeva2023outage, khammassi2023new}, FAS and MIMO are not mutually exclusive~\cite{wang2024fluid}. The integration of these two paradigms, termed MIMO-FAS, can be realized by connecting a massive number of densely packed ports, which enable fine-grained spatial reconfigurability, to a substantially smaller number of RF chains for simultaneous multi-port activation. This dynamic multi-port reconfigurability enables MIMO-FAS to proactively shape the channel matrix into more favorable realizations, yielding significant rate improvements compared to conventional FPA-based MIMO~\cite{new2023information}. In multiuser scenarios, such position flexibility is particularly promising: optimizing the channel matrix corresponds to repositioning the fluid antennas to escape from both deep fades and interference, effectively delivering considerable spatial multiplexing gains without escalating hardware complexity~\cite{zhu2023movable,new2024tutorial}.

\subsection{Prior Work} \label{subsec:prior_work}

Fully unlocking the potential of multiuser MIMO-FAS requires both accurate channel state information~(CSI) acquisition and subsequent port selection (also known as `antenna position optimization'). Fundamentally, the effectiveness of port selection hinges on the quality of the acquired CSI, as estimation errors translate directly into suboptimal port decisions. Despite this inherent coupling, existing literature largely treats these two tasks in isolation~\cite{zhu2023movable, ma2023compressed, xiao2024channel, xu2024sparse, skouroumounis2022fluid, zhang2024successive, chen2025iterative, guo2025channel, xu2023channel, qin2024antenna, cheng2024sum, liao2025joint, cheng2024exploiting, xiao2024multiuser, wu2025globally}: available channel estimators are developed independently of downstream objectives, while most port selection algorithms assume the availability of perfect CSI. In practical deployments, this decoupled treatment inevitably results in a cascading performance degradation.

Channel estimation for FAS is highly challenging: an extremely high-dimensional channel matrix must be recovered from a small number of port observations, and accumulating sufficient observations inevitably incurs a substantial switching cost. Most existing works tackled this problem in SISO scenarios, leveraging either the intrinsic channel sparsity via compressed sensing~(CS)-based methods~\cite{ma2023compressed, xiao2024channel, xu2024sparse} or the spatial correlation among densely deployed FAS ports~\cite{skouroumounis2022fluid, zhang2024successive, chen2025iterative}. However, CS-based approaches suffer from an extremely singular measurement matrix when formulated over high-resolution angular dictionaries, which degrades the estimation accuracy~\cite{guo2025channel}. Moreover, while the switching cost of these approaches might be acceptable for SISO-FAS, it becomes prohibitive when extended to multiuser MIMO-FAS, where the observation burden aggregates across multiple users and severely restricts system scalability. In the multiuser context, a low-sample-size sparse channel recovery (L3SCR) framework was proposed in~\cite{xu2023channel} to specifically tackle FAS-aided multiuser uplink channel estimation. However, this method strictly relies on complete array observations at the BS. \revb{Beyond FAS, channel estimation over dense quasi-continuous apertures has also been investigated for holographic MIMO systems, where the enlarged aperture places users and scatterers in a hybrid near-field and far-field regime and the resulting power diffusion effect calls for dedicated sparse recovery designs~\cite{yue2024hybrid}.}

\revb{In parallel with these model-based approaches, deep generative models have recently been introduced as learned channel priors for conventional MIMO channel estimation. For instance, a score-based prior combined with variational inference was developed in~\cite{chen2025generative}, and a follow-up work simplifies the iterative refinement into a single transformer forward pass under full unitary pilot observations~\cite{chen2026sampling}. Flow matching priors have also been investigated, either embedded in a plug-and-play proximal gradient descent (PnP-PGD) scheme that supports posterior sampling~\cite{zhang2026channel} or driven by a recursive fixed-point solver with anchored trajectory rectification~\cite{jiang2026recursive}. While these works establish the effectiveness of learned generative priors for MIMO channel estimation, their design objective centers on estimation fidelity, since the reconstructed channel itself is the ultimate target. In multiuser MIMO-FAS, however, the channel estimate serves as an intermediate quantity for the subsequent port selection, which must evaluate all candidate ports including those left unobserved due to port-domain sub-sampling. The estimator should therefore convey the uncertainty over the complete channel to the downstream decision, and its efficacy is ultimately determined by the resulting selection quality rather than by reconstruction accuracy alone.}

\revb{On the port selection side,} prior studies have extensively investigated the joint design of antenna positioning and beamforming/combining largely in isolation from the channel estimation process, typically assuming either perfect CSI or imperfect CSI with fixed statistical estimation errors~\cite{zhu2023movable, xiao2024multiuser, wu2025globally}. These works targeted diverse objectives, such as total transmit power minimization~\cite{zhu2023movable, qin2024antenna, cheng2024exploiting}, sum-rate maximization~\cite{cheng2024sum, liao2025joint}, and max-min fairness~\cite{xiao2024multiuser}. Due to the coupled optimization variables and non-convex objectives, existing approaches typically resort to suboptimal solutions, either via alternating optimization (AO) techniques~\cite{qin2024antenna, cheng2024sum, cheng2024exploiting} or heuristic antenna positioning, such as multi-directional descent (MDD)~\cite{zhu2023movable} and particle swarm optimization (PSO)~\cite{xiao2024multiuser}, under fixed beamforming/combining. However, these iterative methods are prone to being trapped in local optima. Although a recent work~\cite{wu2025globally} achieved jointly optimal beamforming and antenna positioning under discrete antenna movements, its prohibitive computational complexity renders it unscalable for high-dimensional FAS. Consequently, the existing literature lacks a unified framework for multiuser MIMO-FAS that jointly accounts for channel estimation and port selection under practical conditions.

\subsection{Contributions}

In this paper, we propose a unified two-stage framework driven by generative diffusion models to systematically address both channel estimation and port selection in multiuser MIMO-FAS scenarios. Unlike conventional decoupled designs, the two stages in our framework are not independently considered. Instead, they are rigorously derived from a single maximum-a-posteriori (MAP) inference problem \revb{whose solution coincides with maximizing the posterior expected utility of the port selection, where the complete channel serves as a latent variable. The two stages approximate this objective with dedicated diffusion models responsible for posterior sampling and candidate proposal, respectively, while the final selection is always governed by the underlying decision criterion, and the conventional cascaded pipeline is regarded as a degenerate special case.} Aligned with realistic hardware implementations, we focus on a two-dimensional (2D) discrete FAS architecture~\cite{pringle2004reconfigurable, zhuravlev2015experimental, zhang2024novel}. The main contributions of this paper are summarized as follows:
\begin{itemize}
    \item First, we establish a comprehensive system model for the uplink multiuser MIMO-FAS system, characterizing the high-dimensional 2D FAS channel via multi-path propagation modeling, the sparse spatial sampling model for pilot observation, and the port selection-dependent effective channel for multiuser data reception. On this basis, we develop a unified probabilistic formulation that casts the joint channel estimation and port selection task as a single MAP inference problem \revb{with the posterior expected utility of the port selection as the decision criterion. Since this objective involves a high-dimensional expectation over the latent channel space and the outer maximization remains combinatorial, we derive the proposed two-stage framework as a tractable approximation, in which an ensemble of posterior channel samples replaces this expectation and the final selection maximizes the resulting empirical criterion, with the single-sample case reducing to the conventional cascaded pipeline.}
    \item For Stage~I, which aims to accurately reconstruct the channel from extremely limited observations, we exploit a flow-based diffusion model~\cite{chen2018neural} trained via the flow matching framework~\cite{lipman2022flow, albergo2022building, liu2022flow} to serve as a powerful learned implicit prior for 2D FAS channels. To enable simultaneous multiuser channel estimation, we develop a guided generation mechanism inspired by the flow-driven posterior sampling (FlowDPS) solver for inverse problems~\cite{kim2025flowdps}, effectively realizing approximate sampling from the channel posterior established in the proposed probabilistic formulation. Numerical experiments show that our proposed estimator achieves remarkable accuracy with only 20 inference steps and maintains exceptional robustness even under severely low sub-sampling ratios.
    \item \revb{For Stage~II, we employ a discrete diffusion model as a learned conditional sampler that proposes a pool of promising port selection candidates conditioned on the estimated channel information from Stage~I. The sampler is trained via supervised learning on optimal labels obtained through exhaustive search, so that its sampling concentrates on high-utility selections, and the final decision maximizes the empirical posterior expected utility over the candidate pool, followed by a lightweight local refinement under the same criterion. Simulation results show that our proposed unified two-stage diffusion framework simultaneously achieves accurate channel recovery and attains over $90\%$ of the deployed rate of the computationally expensive exhaustive search over the estimated channel, which translates into a substantial improvement in the minimum achievable rate relative to practical cascaded baselines.}
\end{itemize}

The remainder of this paper is organized as follows. Section~\ref{sec:sys_model} presents the system model and develops the unified probabilistic formulation that motivates the two-stage decomposition. Section~\ref{sec:chan_est} details the proposed flow-based diffusion framework for Stage~I channel posterior sampling. Section~\ref{sec:port_sel} develops a discrete diffusion-based solver for Stage~II conditional port selection. Finally, Section~\ref{sec:numerical_results} evaluates the performance of the proposed unified framework via numerical simulations, and Section~\ref{sec:conclusion} concludes the paper.

\textit{Notations:} $a$, $\mathbf{a}$, $\mathbf{A}$, and $\mathcal{A}$ represent a scalar, a vector, a matrix, and a set, respectively. $\mathbb{R}^{M \times N}$ and $\mathbb{C}^{M \times N}$ denote the sets of real and complex matrices of dimension $M \times N$, respectively. $(\cdot)^{\mathsf{T}}$, $(\cdot)^{\mathsf{H}}$, $(\cdot)^{-1}$, and $(\cdot)^{\dagger}$ denote the transpose, conjugate transpose, inverse, and Moore-Penrose pseudo-inverse, respectively. $\mathrm{vec}(\mathbf{A})$ denotes the vectorization of matrix $\mathbf{A}$. $\Re(\cdot)$ and $\Im(\cdot)$ denote the real and imaginary parts of a complex variable, respectively. $\otimes$ represents the Kronecker product. $|\mathcal{A}|$ represents the cardinality of set $\mathcal{A}$. $\|\mathbf{a}\|_1$, $\|\mathbf{a}\|_2$, and $\|\mathbf{A}\|_{\mathrm{F}}$ represent the $\ell_1$-norm, $\ell_2$-norm, and Frobenius norm, respectively. $[\mathbf{a}]_{i:j}$ denotes the slicing operation, extracting the $i$-th through $j$-th entries of vector $\mathbf{a}$. $\mathbb{E}[\cdot]$ denotes the expected value of a random variable/vector. $\mathbf{I}_N$ denotes the identity matrix with size $N \times N$. $\mathcal{CN}(\mathbf{a}, \mathbf{A})$ denotes the circularly symmetric complex Gaussian (CSCG) distribution with mean vector $\mathbf{a}$ and covariance matrix $\mathbf{A}$. $D_{\text{KL}}(p \parallel q)$ denotes the Kullback-Leibler (KL) divergence between probability distributions $p$ and $q$.

\section{System Model and Problem Formulation} \label{sec:sys_model}

\begin{figure}[t]
\centering
\includegraphics[width=0.8\linewidth, trim=0cm 0.6cm 0cm 0.6cm]{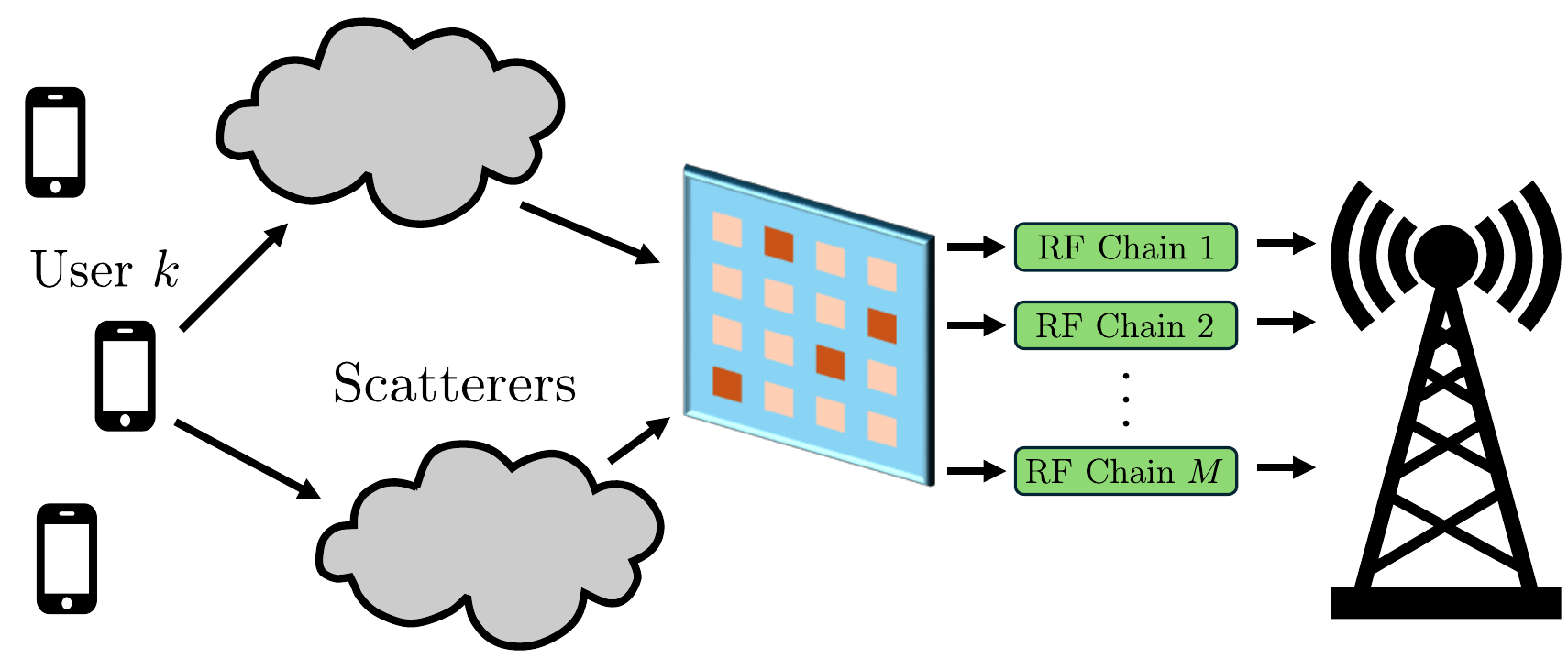}
\caption{An illustration of the multiuser MIMO-FAS system model.}
\label{fig:system_model}
\end{figure}

\subsection{Channel Model}

As illustrated in Fig. \ref{fig:system_model}, we consider an uplink multiuser wireless communication system where a base station (BS) equipped with a 2D FAS serves $K$ single-FPA users. The FAS features a rectangular surface of dimensions $W = W_x\lambda \times W_y\lambda$, with $\lambda$ being the carrier wavelength. This surface is uniformly partitioned into a grid of $N = N_x \times N_y$ discrete candidate ports. To enable spatial multiplexing of the $K$ users, the BS incorporates $M$ RF chains ($N \gg M \ge K$), each connected to a dedicated fluid antenna that can dynamically switch among the $N$ ports.

Let $\mathbf{H}_k \in \mathbb{C}^{N_x \times N_y}$ denote the complete uplink CSI matrix from user $k$ to all FAS ports at the BS. To characterize the multi-path propagation environment, we consider a finite scattering channel model~\cite{new2024tutorial}. Under this model, assuming the wireless channel is composed of $N_p$ dominant paths, the complex channel matrix $\mathbf{H}_k$ can be formulated as
\begin{equation}
    \mathbf{H}_k = \sqrt{\frac{1}{N_p}} \sum_{i=1}^{N_p} g_{k,i} \mathbf{a}_x(\theta_{k,i}, \phi_{k,i}) \mathbf{a}_y(\theta_{k,i}, \phi_{k,i})^{\mathsf{T}},
\end{equation}
where $g_{k,i}$ corresponds to the complex baseband gain of the $i$-th propagation path for user $k$, while $\theta_{k,i}$ and $\phi_{k,i}$ represent the associated elevation and azimuth angles-of-arrival (AoAs), respectively. Furthermore, the receive steering vectors along the horizontal (x-axis) and vertical (y-axis) dimensions, denoted by $\mathbf{a}_x(\theta, \phi)$ and $\mathbf{a}_y(\theta, \phi)$, are respectively given by\footnote{Although the vertical steering vector $\mathbf{a}_y$ depends only on the elevation angle $\theta$, the azimuth angle $\phi$ is still kept in the argument to maintain notational symmetry with $\mathbf{a}_x$.}
\begin{equation}
\begin{aligned}
\mathbf{a}_x(\theta,\phi) &= \left[1, e^{-j 2\pi \frac{W_x}{N_x-1} \cos\theta\sin\phi}, \cdots, e^{-j 2\pi W_x \cos\theta\sin\phi} \right]^T, \\
\mathbf{a}_y(\theta,\phi) &= \left[1, e^{-j 2\pi \frac{W_y}{N_y-1} \sin\theta}, \cdots, e^{-j 2\pi W_y \sin\theta} \right]^T.
\end{aligned}
\end{equation}

\subsection{Signal Model} \label{subsec:signal_model}

The uplink communication is structured into two consecutive phases: a pilot transmission phase for channel observation, followed by a data transmission phase in which the BS selects a subset of $M$ ports for signal reception. \revb{We adopt a quasi-static block-fading model in which the channel remains constant over both phases, consistent with the low-mobility indoor scenario considered in Section~\ref{sec:numerical_results}.}

During the pilot transmission phase, the $K$ users synchronously transmit mutually orthogonal pilot sequences, defined by a codebook matrix $\mathbf{P} \in \mathbb{C}^{K \times K}$ satisfying $\mathbf{P}\mathbf{P}^\mathsf{H} = \mathbf{I}_K$, to prevent inter-user interference. Constrained by the $M$ RF chains, only $M$ ports can be observed simultaneously. To accumulate sufficient spatial observations across the $N$ candidate ports, the pilot block is repeatedly transmitted over $L$ consecutive cycles, with a different subset of $M$ ports activated in each cycle. \revb{A total of $N_{\mathcal{O}}$ port observations are thus collected over $L = \lceil N_{\mathcal{O}}/M \rceil$ cycles,} yielding a sub-sampling ratio of $\delta = N_{\mathcal{O}}/N$. \revb{The resulting pilot overhead amounts to $LK$ symbols, and the choice of $N_{\mathcal{O}}$ directly trades off channel recovery accuracy against pilot overhead, as evaluated in Section~\ref{sec:subsampling_ratio}.} Let $\mathcal{O} \subset \{1,\ldots, N\}$ denote the ordered index set of the sampled ports across all cycles, where $|\mathcal{O}| = N_{\mathcal{O}}$. The overall spatial sampling process can then be represented by a binary selection matrix $\boldsymbol{\Omega}_{\mathcal{O}} \in \{0,1\}^{N_{\mathcal{O}} \times N}$, whose rows are selected from $\mathbf{I}_N$ according to the indices in $\mathcal{O}$.

By stacking the received pilot signals from all $L$ cycles, the aggregated received signal matrix $\mathbf{Y}_\mathrm{p} \in \mathbb{C}^{N_\mathcal{O} \times K}$ at the BS is given by
\begin{equation}
    \mathbf{Y}_\mathrm{p} = \boldsymbol{\Omega}_\mathcal{O} \tilde{\mathbf{H}} \mathbf{P} + \mathbf{N}, \label{eq:observation_model}
\end{equation}
where $\tilde{\mathbf{H}} = [\tilde{\mathbf{h}}_1, \tilde{\mathbf{h}}_2, \dots, \tilde{\mathbf{h}}_K] \in \mathbb{C}^{N \times K}$ represents the concatenated complete CSI matrix for all $K$ users, with $\tilde{\mathbf{h}}_k = \mathrm{vec}(\mathbf{H}_k)$ being the vectorized channel of user $k$. Here, $\mathbf{N} \in \mathbb{C}^{N_\mathcal{O} \times K}$ denotes the received additive white Gaussian noise (AWGN) matrix, whose entries are independent and identically distributed (i.i.d.) according to $\mathcal{CN}(0, \sigma_n^2)$.

To facilitate learning-based signal processing, we define a real-valued equivalent representation for each user's channel, denoted by $\mathbf{h}_k$. Specifically, we introduce an isomorphic mapping $f: \mathbb{C}^{N \times 1} \to \mathbb{R}^{2N \times 1}$ that converts the complex channel vectors to their real-valued counterparts. Accordingly, $\mathbf{h}_k$ can be expressed as
\begin{equation}
    \mathbf{h}_k = f(\tilde{\mathbf{h}}_k) = \begin{bmatrix} \Re(\tilde{\mathbf{h}}_k) \\ \Im(\tilde{\mathbf{h}}_k) \end{bmatrix}. \label{eq:re_equiv_chan}
\end{equation}
The corresponding inverse mapping, denoted by $f^{-1}: \mathbb{R}^{2N \times 1} \to \mathbb{C}^{N \times 1}$, does the opposite by mapping the real-valued network outputs back to the complex domain, i.e.,
\begin{equation}
    \tilde{\mathbf{h}}_k = f^{-1}(\mathbf{h}_k) = [\mathbf{h}_k]_{1:N} + j [\mathbf{h}_k]_{N+1:2N}.
\end{equation}
As a result, the complete multiuser channel matrix $\tilde{\mathbf{H}}$ can be explicitly written in terms of the real-valued individual channel vectors as
\begin{equation}
    \tilde{\mathbf{H}} = [f^{-1}(\mathbf{h}_1), f^{-1}(\mathbf{h}_2), \dots, f^{-1}(\mathbf{h}_K)]. \label{eq:H_tilde_reconstruction}
\end{equation}

In the data transmission phase, the BS activates $M$ out of the $N$ candidate ports for uplink data reception. Mathematically, we represent the port selection decision by a binary mask $\mathbf{x} \in \{0,1\}^N$ with $\|\mathbf{x}\|_1 = M$, where $x_i = 1$ indicates that the $i$-th port is activated, and $x_i = 0$ otherwise. The corresponding port selection matrix $\boldsymbol{\Omega}_\mathbf{x} \in \{0,1\}^{M \times N}$ is formed by extracting the rows of $\mathbf{I}_N$ indexed by the nonzero entries of $\mathbf{x}$, which yields the effective multiuser channel
\begin{equation}
    \bar{\mathbf{H}} = \boldsymbol{\Omega}_\mathbf{x} \tilde{\mathbf{H}} \in \mathbb{C}^{M \times K}. \label{eq:effective_channel}
\end{equation}

The baseband received signal at the BS is then given by
\begin{equation}
    \mathbf{y}_\mathrm{d} = \bar{\mathbf{H}} \mathbf{s} + \mathbf{n}_\mathrm{d},
\end{equation}
where $\mathbf{s} \in \mathbb{C}^{K \times 1}$ represents the transmitted data symbol vector from all $K$ users with normalized signal power $\mathbb{E}[\mathbf{s}\mathbf{s}^{\mathsf{H}}] = \mathbf{I}_K$, and $\mathbf{n}_\mathrm{d} \sim \mathcal{CN}(\mathbf{0}, \sigma_n^2 \mathbf{I}_M)$ is the AWGN vector.

\subsection{Problem Formulation} \label{subsec:problem_formulation}

For uplink signal detection, a linear receive combining scheme is adopted at the BS. With a combining matrix $\mathbf{W} \in \mathbb{C}^{M \times K}$, the decoded data vector is given by $\hat{\mathbf{s}} = \mathbf{W}^{\mathsf{H}}\mathbf{y}_\mathrm{d}$. To simplify our analysis and focus purely on the port selection design, we adopt the minimum mean squared error (MMSE) receiver~\cite{tse2005fundamentals}. \revb{This choice incurs no loss of optimality within the class of linear receivers, since for any given port selection and channel realization the MMSE combiner simultaneously maximizes the receive SINR of every user~\cite{tse2005fundamentals}.} Given perfect knowledge of the complete channel $\tilde{\mathbf{H}}$, the MMSE combining matrix is given by
\begin{equation}
    \mathbf{W}_{\mathrm{MMSE}}(\mathbf{x}, \tilde{\mathbf{H}}) = \left( \boldsymbol{\Omega}_\mathbf{x} \tilde{\mathbf{H}} \tilde{\mathbf{H}}^\mathsf{H} \boldsymbol{\Omega}_\mathbf{x}^\mathsf{T} + \sigma_n^2 \mathbf{I}_M \right)^{-1} \boldsymbol{\Omega}_\mathbf{x} \tilde{\mathbf{H}}. \label{eq:mmse_receiver}
\end{equation}
Let $\mathbf{w}_k(\mathbf{x}, \tilde{\mathbf{H}})$ and $\bar{\mathbf{h}}_k$ denote the $k$-th columns of $\mathbf{W}_{\mathrm{MMSE}}(\mathbf{x}, \tilde{\mathbf{H}})$ and the true effective channel $\bar{\mathbf{H}}$, respectively. The corresponding receive signal-to-interference-plus-noise ratio (SINR) for user $k$ is expressed as
\begin{equation}
    \gamma_k(\mathbf{x}, \tilde{\mathbf{H}}) = \frac{|\mathbf{w}_k(\mathbf{x}, \tilde{\mathbf{H}})^\mathsf{H} \bar{\mathbf{h}}_k|^2}{\sum_{j \neq k} |\mathbf{w}_k(\mathbf{x}, \tilde{\mathbf{H}})^\mathsf{H} \bar{\mathbf{h}}_j|^2 + \sigma_n^2 \|\mathbf{w}_k(\mathbf{x}, \tilde{\mathbf{H}})\|_2^2}, \label{eq:sinr}
\end{equation}
with the achievable rate for user $k$ given by
\begin{equation}
    R_k(\mathbf{x}, \tilde{\mathbf{H}}) = \log_2(1 + \gamma_k(\mathbf{x}, \tilde{\mathbf{H}})). \label{eq:rate}
\end{equation}

To ensure performance fairness among all active users, our objective is to optimize the port selection strategy $\mathbf{x}$ by maximizing the minimum achievable rate, subject to FAS hardware constraints. The corresponding max-min optimization problem can be formulated as
\begin{subequations}\label{eq:port_sel_orig}
\begin{align}
\max_{\mathbf{x}} \quad & \min_{k \in \{1,\dots,K\}} R_k(\mathbf{x}, \tilde{\mathbf{H}}) \label{eq:port_sel_orig_a} \\
\text{s.t.} \quad & \mathbf{x} \in \{0, 1\}^N, \label{eq:port_sel_orig_b} \\
& \|\mathbf{x}\|_1 = M. \label{eq:port_sel_orig_c}
\end{align}
\end{subequations}
Directly solving problem~\eqref{eq:port_sel_orig} in practice is intractable due to two fundamental challenges. First, the objective relies on the complete true channel $\tilde{\mathbf{H}}$, which is inaccessible at the BS since only the highly sub-sampled pilot signals $\mathbf{Y}_\mathrm{p}$ are observed. Second, even if $\tilde{\mathbf{H}}$ were perfectly known, optimizing the discrete binary mask $\mathbf{x}$ poses a \revb{combinatorial optimization (CO) problem whose feasible set contains $\binom{N}{M}$ candidate subsets, a size that grows exponentially with the number of selected ports. Finding the optimal solution via online exhaustive search therefore incurs a prohibitive computational burden for large-scale FAS.}

To circumvent these challenges, we adopt a Bayesian perspective and treat the unknown $\tilde{\mathbf{H}}$ as a latent random variable. \revb{Given the observations $\mathbf{Y}_\mathrm{p}$, the uncertainty over $\tilde{\mathbf{H}}$ is characterized by the channel posterior $p(\tilde{\mathbf{H}}|\mathbf{Y}_\mathrm{p}) \propto p(\mathbf{Y}_\mathrm{p}|\tilde{\mathbf{H}})\,q(\tilde{\mathbf{H}})$, where the likelihood $p(\mathbf{Y}_\mathrm{p}|\tilde{\mathbf{H}})$ is defined by the observation model in~\eqref{eq:observation_model} and the prior factorizes as $q(\tilde{\mathbf{H}}) = \prod_{k=1}^K q(\mathbf{h}_k)$ assuming spatial independence among geographically distributed users. Under such channel uncertainty, a natural decision criterion is to maximize the \emph{posterior expected utility}
\begin{equation}
    \bar{U}(\mathbf{x}; \mathbf{Y}_\mathrm{p}) = \mathbb{E}_{\tilde{\mathbf{H}} \sim p(\tilde{\mathbf{H}}|\mathbf{Y}_\mathrm{p})}\big[ U(\mathbf{x}, \tilde{\mathbf{H}}) \big], \label{eq:post_exp_utility}
\end{equation}
where $U(\mathbf{x}, \tilde{\mathbf{H}}) = \min_{k} R_k(\mathbf{x}, \tilde{\mathbf{H}})$ is the utility function corresponding to problem~\eqref{eq:port_sel_orig}. To embed this criterion into a probabilistic framework, we define a Gibbs distribution $p_\kappa(\mathbf{x}|\mathbf{Y}_\mathrm{p}) \propto \exp\big(\kappa \cdot \bar{U}(\mathbf{x}; \mathbf{Y}_\mathrm{p})\big)$ over the feasible set, where $\kappa > 0$ serves as the inverse temperature scale. The overall system objective then translates into a maximum-a-posteriori (MAP) estimation problem:
\begin{equation}
    \mathbf{x}^* = \arg\max_{\mathbf{x} \in \mathcal{X}} \; \revb{p_\kappa(\mathbf{x}|\mathbf{Y}_\mathrm{p}) = \arg\max_{\mathbf{x} \in \mathcal{X}} \; \bar{U}(\mathbf{x}; \mathbf{Y}_\mathrm{p})}, \label{eq:map_objective}
\end{equation}
where $\mathcal{X} = \{\mathbf{x} \in \{0, 1\}^N : \|\mathbf{x}\|_1 = M\}$ represents the feasible set of all valid port selection strategies. The second equality in~\eqref{eq:map_objective} arises from the monotonicity of the exponential function for any $\kappa > 0$. The decision criterion is therefore independent of the specific temperature value, and $\kappa$ only determines the concentration sharpness of $p_\kappa$ around the maximizer of the posterior expected utility.
}

Directly solving the MAP problem in~\eqref{eq:map_objective} still remains intractable \revb{since the posterior expected utility $\bar{U}(\mathbf{x};\mathbf{Y}_\mathrm{p})$ involves a high-dimensional expectation over the entire latent channel space and the outer maximization retains the combinatorial structure of the feasible set $\mathcal{X}$. Our framework therefore approximates~\eqref{eq:map_objective} through two sequential stages. Stage~I addresses the channel estimation task, which aims to infer the unknown channel $\tilde{\mathbf{H}}$ from the sub-sampled pilot observations $\mathbf{Y}_\mathrm{p}$. For this purpose,} we employ a flow-based diffusion model that learns a strong data-driven structural prior over the channel and performs approximate posterior sampling via measurement-guided generation, as detailed in Section~\ref{sec:chan_est}. \revb{This stage yields an ensemble of $S$ posterior channel samples, denoted by $\{\hat{\mathbf{H}}^{(s)}\}_{s=1}^S$, each consistent with the pilot observations $\mathbf{Y}_\mathrm{p}$ while respecting the learned prior. The intractable decision criterion in~\eqref{eq:post_exp_utility} is in turn approximated by its empirical counterpart
\begin{equation}
    \hat{\bar{U}}_S(\mathbf{x}) = \frac{1}{S} \sum_{s=1}^{S} U\big(\mathbf{x}, \hat{\mathbf{H}}^{(s)}\big), \label{eq:empirical_utility}
\end{equation}
which serves as an increasingly accurate approximation of $\bar{U}(\mathbf{x};\mathbf{Y}_\mathrm{p})$ as $S$ grows.
}

\begin{figure}[!t]
    \centering
    \includegraphics[width=0.85\linewidth]{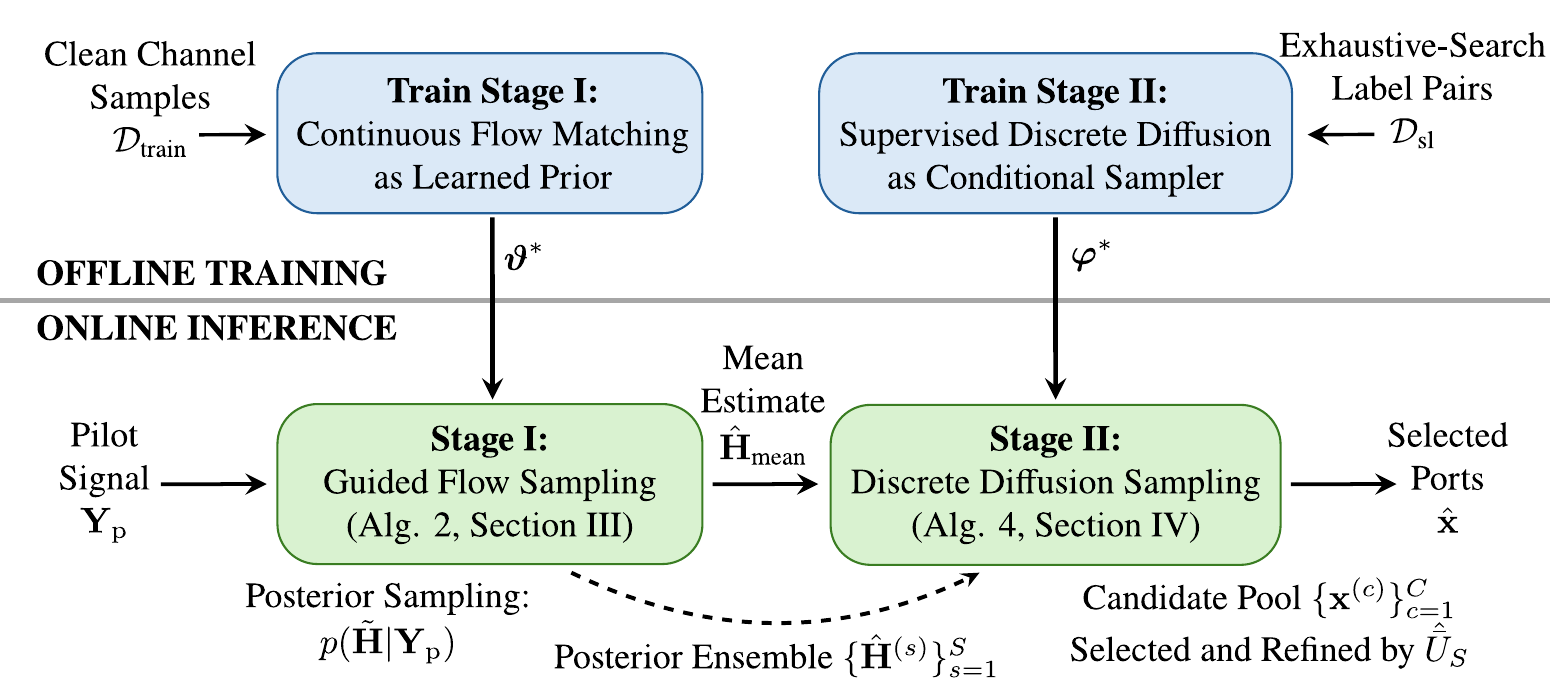}
    \caption{Overall workflow of the proposed two-stage diffusion framework.}
    \label{fig:overall_workflow}
\end{figure}

\revb{Stage~II subsequently addresses the port selection task, which amounts to maximizing~\eqref{eq:empirical_utility} over the combinatorial feasible set $\mathcal{X}$, whose exhaustive enumeration is infeasible online. Instead of searching $\mathcal{X}$ directly, we train a discrete diffusion model as a learned conditional sampler that proposes a pool of $C$ promising port selection candidates $\{\mathbf{x}^{(c)}\}_{c=1}^{C}$ conditioned on the estimated channel information from Stage~I, as detailed in Section~\ref{sec:port_sel}. The final decision then maximizes the empirical criterion over this pool,
\begin{equation}
    \hat{\mathbf{x}} = \arg\max_{\mathbf{x}\in\{\mathbf{x}^{(c)}\}_{c=1}^{C}} \hat{\bar{U}}_S(\mathbf{x}), \label{eq:pool_decision}
\end{equation}
followed by a lightweight local refinement under the same criterion. In this way, generative sampling is responsible only for proposing promising candidates, whereas the selection among them is always performed by the decision criterion itself, so that the two stages jointly constitute a tractable approximation of the unified MAP problem in~\eqref{eq:map_objective}.} The overall workflow of the proposed framework is illustrated in Fig.~\ref{fig:overall_workflow}.

\revb{Notably, the special case $S = 1$ reduces~\eqref{eq:empirical_utility} to the} single-sample Monte Carlo (plug-in) approximation, which corresponds to replacing the channel posterior with a Dirac delta function centered at a single posterior sample $\hat{\mathbf{H}}$, i.e., $p(\tilde{\mathbf{H}}|\mathbf{Y}_\mathrm{p}) \approx \delta(\tilde{\mathbf{H}} - \hat{\mathbf{H}})$. \revb{The high-dimensional expectation in~\eqref{eq:post_exp_utility} thereby collapses into a point evaluation, where $\hat{\mathbf{H}}$ acts as a deterministic proxy for the true channel and the port selection is optimized solely against this proxy, which is precisely the pipeline of conventional cascaded designs.}

\begin{remark}[Unified MAP versus Heuristic Cascade]
Although the resulting sequential execution resembles the conventional cascaded pipeline, our proposed two-stage framework is derived as a principled approximation of the unified MAP objective in~\eqref{eq:map_objective}, which provides a rigorous mathematical justification for the sequential design. \revb{More importantly, the unified formulation induces a decision mechanism that a conventional cascade does not possess. In particular, the decision rule in~\eqref{eq:pool_decision} scores candidates against the entire ensemble of posterior samples according to~\eqref{eq:empirical_utility}, whereas a conventional cascaded design evaluates candidates only through a single deterministic estimation proxy. Any point estimator is inherently restricted to such single-estimate evaluation, since it provides only a point estimate instead of the channel posterior.}
\end{remark}

\section{Flow-Based Diffusion Framework for Channel Estimation} \label{sec:chan_est}

As illustrated in Section~\ref{subsec:problem_formulation}, Stage~I requires approximating and sampling from the channel posterior $p(\tilde{\mathbf{H}}|\mathbf{Y}_\mathrm{p})$. Hand-crafted priors, such as the angular-domain sparsity assumed in~\cite{ma2023compressed, xiao2024channel, xu2024sparse}, fail to capture the intricate spatial correlations of FAS channels, making them inadequate to regularize the severely under-determined inverse problem in~\eqref{eq:observation_model}. To bridge this gap, we adopt a flow-based diffusion model to learn a powerful data-driven channel prior, whose generation process naturally accommodates measurement-guided approximate posterior sampling for channel estimation.

In this section, we first briefly outline the fundamentals of flow-based diffusion models, then introduce our offline prior learning strategy, and finally elaborate on the proposed online channel estimation framework via parallel guided generation.

\subsection{Preliminaries on Flow-Based Diffusion Models} \label{subsec:preliminaries_on_flow}

The fundamental objective of generative modeling is to learn a parameterized model $p_{\boldsymbol{\vartheta}}(\mathbf{h})$ that approximates a true data distribution $q(\mathbf{h})$, given a set of training samples $\mathcal{D}_\text{train} = \{\mathbf{h}^{(i)}\}_{i=1}^{N_{\text{train}}}$ drawn from it\footnote{We omit the user index $k$ for notational simplicity and use $\mathbf{h}$ to denote the real-equivalent channel vector of an individual user, defined in~\eqref{eq:re_equiv_chan}.}. Flow-based diffusion models achieve this by learning a continuous transformation from a standard Gaussian source distribution $\mathcal{N}(0,\mathbf{I})$ to the target distribution $q(\mathbf{h})$ via a continuous-time probability path $\{p_t\}_{t\in[0,1]}$~\cite{chen2018neural, lipman2022flow, albergo2022building, liu2022flow}, where $t \in [0,1]$ parameterizes the evolution such that $p_1 = \mathcal{N}(0,\mathbf{I})$ and $p_0 = q$.

Mathematically, this transformation is characterized by a time-dependent bijective mapping $\psi_t$, referred to as a \textit{flow}, governed by an ordinary differential equation (ODE):
\begin{equation}
    \frac{\mathrm{d}\psi_t(\mathbf{z})}{\mathrm{d}t} = \mathbf{v}_t(\psi_t(\mathbf{z})), \quad \text{with} \quad \psi_1(\mathbf{z}) = \mathbf{z}, \label{eq:flow_def}
\end{equation}
where $\mathbf{v}_t$ denotes the time-dependent velocity field. Physically, $\mathbf{v}_t$ drives an initial noise sample $\mathbf{z}_1 \sim \mathcal{N}(\mathbf{0},\mathbf{I})$ along a continuous trajectory $\mathbf{z}_t = \psi_t(\mathbf{z}_1)$ such that its marginal distribution precisely matches $p_t$ throughout the evolution. Therefore, by solving the ODE from $t=1$ to $t=0$, the flow deterministically maps $\mathbf{z}_1$ into a structured channel realization $\mathbf{h} = \mathbf{z}_0 = \psi_0(\mathbf{z}_1)$ that follows the target distribution.

To train this generative model, suppose we have defined a target marginal probability path $p_t$ connecting the source distribution $p_1 = p$ to the target data distribution $p_0 = q$, and let $\mathbf{v}_t$ be its corresponding marginal velocity field. We can then approximate $\mathbf{v}_t$ using a neural network $\mathbf{v}_t^{\boldsymbol{\vartheta}}$ via the Flow Matching (FM) objective:
\begin{equation}
    \mathcal{L}_{\text{FM}}(\boldsymbol{\vartheta}) = \mathbb{E}_{t, \mathbf{z}_t \sim p_t} \left[ \left\| \mathbf{v}_t(\mathbf{z}_t) - \mathbf{v}_t^{\boldsymbol{\vartheta}}(\mathbf{z}_t) \right\|_2^2 \right].
\end{equation}

Despite being conceptually straightforward, this objective is intractable to evaluate in practice. Specifically, as suggested by~\cite{lipman2022flow}, a feasible way to construct $p_t$ is through marginalization over simpler conditional probability paths $p_t(\mathbf{z}_t|\mathbf{h})$:
\begin{equation}
    p_t(\mathbf{z}_t) = \int p_t(\mathbf{z}_t|\mathbf{h})q(\mathbf{h})\mathrm{d}\mathbf{h}.
\end{equation}
As a result, the corresponding marginal velocity field $\mathbf{v}_t$ can be expressed as the posterior expectation of the conditional velocity fields:
\begin{equation}
    \begin{aligned}
\mathbf{v}_t(\mathbf{z}_t) &= \mathbb{E}_{\mathbf{h} \sim p_t(\cdot|\mathbf{z}_t)}[\mathbf{v}_t(\mathbf{z}_t | \mathbf{h})] \\
&= \int \mathbf{v}_t(\mathbf{z}_t | \mathbf{h}) \frac{p_t(\mathbf{z}_t | \mathbf{h}) q(\mathbf{h})}{p_t(\mathbf{z}_t)} \mathrm{d}\mathbf{h},
\end{aligned} \label{eq:cond_velocity}
\end{equation}
where $\mathbf{v}_t(\mathbf{z}_t | \mathbf{h})$ denotes the conditional velocity field that generates $p_t(\mathbf{z}_t|\mathbf{h})$. Evidently, computing this expectation requires intractable integration over the entire unknown data distribution $q(\mathbf{h})$.

To bypass this computational bottleneck, an alternative Conditional Flow Matching (CFM) objective was proposed in~\cite{lipman2022flow}, which directly regresses the neural network against the conditional velocity field:
\begin{equation}
    \mathcal{L}_{\text{CFM}}(\boldsymbol{\vartheta}) = \mathbb{E}_{t, \mathbf{h} \sim q, \mathbf{z}_t \sim p_t(\cdot|\mathbf{h})} \left[ \left\| \mathbf{v}_t(\mathbf{z}_t|\mathbf{h}) - \mathbf{v}_t^{\boldsymbol{\vartheta}}(\mathbf{z}_t) \right\|_2^2 \right], \label{eq:cfm_objective}
\end{equation}
and it has been proven that optimizing this CFM objective is mathematically equivalent to the original FM objective, since they share the same gradient with respect to $\boldsymbol{\vartheta}$.

Once the network $\mathbf{v}_t^{\boldsymbol{\vartheta}}$ is trained, generation is realized by drawing $\mathbf{z}_1 \sim \mathcal{N}(0, \mathbf{I})$ and then numerically solving the flow ODE from $t=1$ to $t=0$ using solvers like the Euler method:
\begin{equation}
    \mathbf{z}_{t-\Delta t} = \mathbf{z}_t - \mathbf{v}_t(\mathbf{z}_t) \Delta t, \label{eq:euler_solver}
\end{equation}
where $\Delta t$ is the discretization step size. By adjusting the number of function evaluations (NFEs), one can flexibly trade off sample generation quality against inference speed.

\subsection{Offline Training: Learning the Flow-Based Channel Prior}

To instantiate the CFM objective defined in~\eqref{eq:cfm_objective} for model training, we must design a specific conditional probability path $p_t(\mathbf{z}_t|\mathbf{h})$ and its corresponding conditional velocity field $\mathbf{v}_t(\mathbf{z}_t|\mathbf{h})$.

\subsubsection{Affine Conditional Flow and Optimal Transport}

For mathematical simplicity, we consider the affine conditional flow, where the intermediate state $\mathbf{z}_t$ is explicitly constructed by linearly interpolating between a data sample $\mathbf{h} \sim q$ and a specific noise sample $\mathbf{z}_1 \sim \mathcal{N}(0,\mathbf{I})$:
\begin{equation}
    \mathbf{z}_t = \psi_t(\mathbf{z}_1|\mathbf{h}) = a_t \mathbf{h} + b_t \mathbf{z}_1. \label{eq:affine_cond_flow_trajectory}
\end{equation}
To fulfill the required boundary conditions (i.e., $p_0=q$ and $p_1=p$), the time-dependent coefficients must satisfy $a_0 = 1, b_0 = 0$ at $t=0$, and $a_1 = 0, b_1 = 1$ at $t=1$. While any coefficient schedule satisfying these boundary conditions is theoretically valid, we adopt the Optimal Transport (OT) formulation by setting $a_t = 1-t$ and $b_t = t$, i.e.,
\begin{equation}
    \mathbf{z}_t = (1-t)\mathbf{h} + t\mathbf{z}_1. \label{eq:ot_path}
\end{equation}
This OT path forms a straight-line trajectory between any noise-data pair, which minimizes the discretization error of numerical ODE solvers at larger generation step sizes.

Taking the time derivative of~\eqref{eq:affine_cond_flow_trajectory}, the conditional velocity field $\mathbf{v}_t(\mathbf{z}_t|\mathbf{h})$ under the OT path admits a closed form
\begin{equation}
     \mathbf{v}_t(\mathbf{z}_t|\mathbf{h}) = \frac{\mathrm{d}a_t}{\mathrm{d}t} \mathbf{h} + \frac{\mathrm{d}b_t}{\mathrm{d}t} \mathbf{z}_1 = \mathbf{z}_1 - \mathbf{h}. \label{eq:ot_cond_velocity}
\end{equation}

Furthermore, since the randomness of $\mathbf{z}_t$ conditioned on $\mathbf{h}$ originates entirely from the initial noise $\mathbf{z}_1$, we can apply the reparameterization trick to replace the expectation over $\mathbf{z}_t \sim p_t(\cdot|\mathbf{h})$ with an expectation over the noise source $\mathbf{z}_1 \sim p$. Thus, the CFM objective for our flow-based model reduces to:
\begin{equation}
\mathcal{L}_{\text{CFM}}(\boldsymbol{\vartheta}) = \mathbb{E}_{t, \mathbf{h}\sim q, \mathbf{z}_1 \sim p} \left[ \left\| \mathbf{v}_t^{\boldsymbol{\vartheta}}(\mathbf{z}_t) - (\mathbf{z}_1 - \mathbf{h}) \right\|_2^2 \right]. \label{eq:ot_cfm_objective}
\end{equation}
This objective is highly tractable, requiring only available training samples, standard Gaussian noise, and basic algebraic operations, thereby completely bypassing any complicated numerical integration.

\subsubsection{Network Architecture and Offline Training} \label{sec:net_architecture}

\begin{figure}[t]
\centering
\includegraphics[width=0.8\linewidth, trim=0cm 0.6cm 0cm 0.6cm]{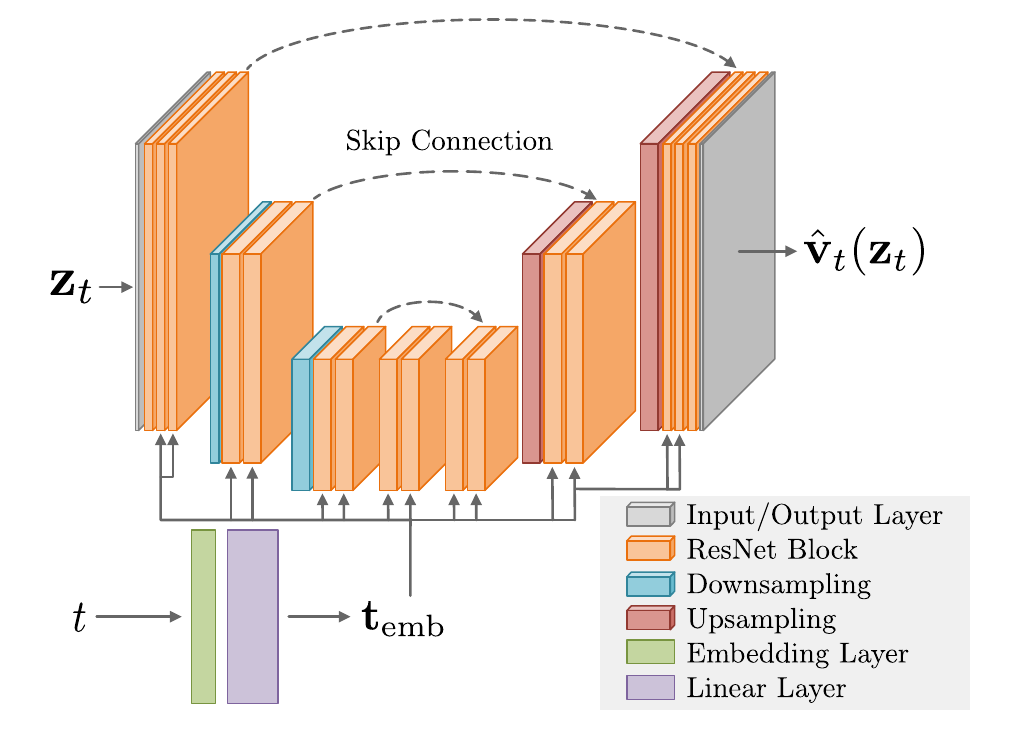}
\caption{An illustration of the velocity network $\mathbf{v}_t^{\boldsymbol{\vartheta}}$, implemented using a lightweight U-Net architecture with sinusoidal time embedding.}
\label{fig:nn_architecture}
\end{figure}

Similar to our previous SISO-FAS work~\cite{tang2025accurate}, we adopt a lightweight U-Net as the backbone of the velocity network $\mathbf{v}_t^{\boldsymbol{\vartheta}}$ to predict $\hat{\mathbf{v}}_t(\mathbf{z}_t)$. As illustrated in Fig.~\ref{fig:nn_architecture}, the network employs a symmetric encoder-decoder topology with three resolution levels, utilizing ResNet blocks for feature extraction and skip connections to preserve fine-grained spatial features. To explicitly exploit the local spatial correlation inherent in 2D FAS channels, the 1D latent variable $\mathbf{z}_t \in \mathbb{R}^{2N \times 1}$ is reshaped into a $2 \times N_x \times N_y$ spatial tensor, where the two channels correspond to the real and imaginary components. Furthermore, to ensure the network is time-aware, the continuous time variable $t \in [0,1]$ is encoded via a sinusoidal positional embedding $\mathbf{t}_{\text{emb}} \in \mathbb{R}^{D_{\text{emb}}}$ and injected into every ResNet block.

We train $\mathbf{v}_t^{\boldsymbol{\vartheta}}$ by minimizing the OT-based CFM objective in~\eqref{eq:ot_cfm_objective} via mini-batch stochastic gradient descent (SGD). At each iteration, given a mini-batch of $B$ samples drawn from the offline dataset $\mathcal{D}_{\text{train}} = \{\mathbf{h}^{(i)}\}_{i=1}^{N_{\text{train}}}$, we independently sample continuous time steps $t^{(i)} \sim \mathcal{U}[0, 1]$ and source noise $\mathbf{z}_1^{(i)} \sim \mathcal{N}(\mathbf{0}, \mathbf{I})$. Following the OT conditional flow formulation, the intermediate state is constructed as $\mathbf{z}_{t^{(i)}}^{(i)} = (1-t^{(i)})\mathbf{h}^{(i)} + t^{(i)}\mathbf{z}_1^{(i)}$. Consequently, the neural network parameters $\boldsymbol{\vartheta}$ are updated using the gradient of the following empirical loss:
\begin{equation}
    \hat{\mathcal{L}}_{\text{CFM}}(\boldsymbol{\vartheta}) = \frac{1}{B} \sum_{i=1}^{B} \left\| \mathbf{v}_{t^{(i)}}^{\boldsymbol{\vartheta}}(\mathbf{z}_{t^{(i)}}^{(i)}) - (\mathbf{z}_1^{(i)} - \mathbf{h}^{(i)}) \right\|_2^2. \label{eq:mini_batch_loss}
\end{equation}
The overall training procedure is summarized in Algorithm~\ref{alg:flow_model_training}.

\begin{algorithm}[t]
\caption{Offline Flow-Based Diffusion Model Training}
\label{alg:flow_model_training}
\begin{algorithmic}[1]
\REQUIRE Training dataset $\mathcal{D}_{\text{train}} = \{\mathbf{h}^{(i)}\}_{i=1}^{N_{\text{train}}}$, batch size $B$, learning rate $\zeta$.
\REPEAT
    \FOR{$i = 1$ \TO $B$}
        \STATE Sample $\mathbf{h}^{(i)}$ from $\mathcal{D}_{\text{train}}$
        \STATE Sample $t^{(i)} \sim \mathcal{U}[0, 1]$
        \STATE Sample $\mathbf{z}_1^{(i)} \sim \mathcal{N}(\mathbf{0}, \mathbf{I})$
        \STATE Construct $\mathbf{z}_{t^{(i)}}^{(i)} = (1-t^{(i)})\mathbf{h}^{(i)} + t^{(i)}\mathbf{z}_1^{(i)}$
    \ENDFOR
    \STATE Compute $\hat{\mathcal{L}}_{\text{CFM}}$ according to Eq.~\eqref{eq:mini_batch_loss}
    \STATE Update $\boldsymbol{\vartheta} \gets \boldsymbol{\vartheta} - \zeta \nabla_{\boldsymbol{\vartheta}} \hat{\mathcal{L}}_{\text{CFM}}$ via SGD/Adam
\UNTIL{convergence}
\end{algorithmic}
\end{algorithm}

\subsection{Online Inference: Simultaneous Multiuser Channel Estimation via Measurement-Consistent Flow Guidance}

When deploying the offline-trained velocity network $\mathbf{v}_t^{\boldsymbol{\vartheta}}$ for online multiuser channel estimation, directly solving the flow ODEs from $K$ independent Gaussian initializations $\mathbf{z}_{k,1} \sim \mathcal{N}(\mathbf{0}, \mathbf{I})$ would yield unconditional random samples that ignore the pilot observations in~\eqref{eq:observation_model}. To enforce measurement consistency during inference, the generative trajectories must be guided to align with the received signals. To this end, we draw inspiration from the FlowDPS~\cite{kim2025flowdps} framework to design a unified measurement consistency scheme that facilitates parallel guided generation across all users, as detailed below.

\subsubsection{Flow Decomposition}

To effectively guide the generation process, it is necessary to isolate the model's instantaneous estimate of the clean channel at any intermediate state. As the generation process is governed by the numerical integration of the flow ODE in~\eqref{eq:euler_solver}, we proceed by decomposing its two core components: the velocity field $\mathbf{v}_t$ and the intermediate state $\mathbf{z}_t$. For brevity, the user index $k$ is omitted in the following derivations.

First, according to~\eqref{eq:cond_velocity}, the instantaneous marginal velocity at a specific intermediate state $\mathbf{z}_t$ is essentially the posterior expectation of the conditional velocities across all trajectories passing through $\mathbf{z}_t$. By substituting the OT conditional velocity~\eqref{eq:ot_cond_velocity} into~\eqref{eq:cond_velocity} and leveraging the linearity of expectation, $\mathbf{v}_t(\mathbf{z}_t)$ can be analytically decomposed as
\begin{equation}
\begin{aligned}
\mathbf{v}_t(\mathbf{z}_t) &= \mathbb{E}[\mathbf{v}_t(\mathbf{z}_t|\mathbf{h})|\mathbf{z}_t] \\
&= \mathbb{E}[\mathbf{z}_1 - \mathbf{h}|\mathbf{z}_t] = \hat{\mathbf{z}}_{1|t} - \hat{\mathbf{h}}_{0|t}, \label{eq:vt_decomp}
\end{aligned}
\end{equation}
where $\hat{\mathbf{h}}_{0|t} = \mathbb{E}[\mathbf{h}|\mathbf{z}_t]$ and $\hat{\mathbf{z}}_{1|t} = \mathbb{E}[\mathbf{z}_1|\mathbf{z}_t]$ denote the MMSE estimates of the clean channel and the corresponding source noise, respectively, given $\mathbf{z}_t$. Similarly, $\mathbf{z}_t$ itself can be decomposed as a weighted average of these two estimates\footnote{With a slight abuse of notation in $\mathbb{E}[ \mathbf{z}_t | \mathbf{z}_t]$, the $\mathbf{z}_t$ in the conditioning represents a specific deterministic realization, while the other $\mathbf{z}_t$ denotes the random variable itself.}:
\begin{equation}
\begin{aligned}
    \mathbf{z}_t &= \mathbb{E}[ \mathbf{z}_t | \mathbf{z}_t] \\
    &= \mathbb{E}[(1-t)\mathbf{h} + t\mathbf{z}_1 | \mathbf{z}_t] = (1-t)\hat{\mathbf{h}}_{0|t} + t\hat{\mathbf{z}}_{1|t}. \label{eq:zt_decomp}
\end{aligned}
\end{equation}

By rearranging~\eqref{eq:vt_decomp} and~\eqref{eq:zt_decomp} to isolate the two MMSE estimates, and replacing the exact marginal velocity with our neural approximation $\mathbf{v}_t^{\boldsymbol{\vartheta}}(\mathbf{z}_t) \approx \mathbf{v}_t(\mathbf{z}_t)$, we obtain the flow-based version of Tweedie's formula under the OT path~\cite{kim2025flowdps}:
\begin{align}
\hat{\mathbf{h}}_{0|t} &= \mathbf{z}_t - t \mathbf{v}_t^{\boldsymbol{\vartheta}}(\mathbf{z}_t), \label{eq:h_hat} \\
\hat{\mathbf{z}}_{1|t} &= \mathbf{z}_t + (1-t)\mathbf{v}_t^{\boldsymbol{\vartheta}}(\mathbf{z}_t), \label{eq:z_hat}
\end{align}
which provides a tractable way to ``peek'' at the clean sample $\hat{\mathbf{h}}_{0|t}$ and the source noise $\hat{\mathbf{z}}_{1|t}$ from any $\mathbf{z}_t$ during generation.

Finally, by substituting the decompositions~\eqref{eq:zt_decomp} and~\eqref{eq:vt_decomp} back into the standard Euler update step~\eqref{eq:euler_solver}, the transition to the next intermediate state $\mathbf{z}_{t-\Delta t}$ can be reformulated as:
\begin{equation}
\begin{aligned}
\mathbf{z}_{t-\Delta t} &= \mathbf{z}_t - \mathbf{v}_t^{\boldsymbol{\vartheta}}(\mathbf{z}_t)\Delta t \\
&= \left[ (1-t)\hat{\mathbf{h}}_{0|t} + t\hat{\mathbf{z}}_{1|t} \right] - \left[ \hat{\mathbf{z}}_{1|t} - \hat{\mathbf{h}}_{0|t} \right]\Delta t \\
&= (1 - (t-\Delta t))\hat{\mathbf{h}}_{0|t} + (t-\Delta t)\hat{\mathbf{z}}_{1|t}. \label{eq:decomposed_euler}
\end{aligned}
\end{equation}
Notably, this decomposed update explicitly reveals that $\mathbf{z}_{t-\Delta t}$ retains the same weighted-combination structure as $\mathbf{z}_t$, with the time coefficients simply updated to $1-(t-\Delta t)$ and $t-\Delta t$. This decoupled formulation provides a natural and tractable anchor to enforce measurement consistency on $\hat{\mathbf{h}}_{0|t}$ at each integration step, as detailed in the subsequent discussion.

\subsubsection{\texorpdfstring{Unified Guidance on $\hat{\mathbf{h}}_{0|t}$}{Joint Guidance on h hat 0|t}}

We now re-introduce the user index $k$ to elaborate on the unified measurement consistency guidance procedure. Specifically, we let $\hat{\mathbf{h}}_{k, 0|t}$ represent the predicted clean channel estimate for every user $k \in \{1, \ldots, K\}$ from the intermediate states $\mathbf{z}_{k,t}$ at time step $t$. Analogous to~\eqref{eq:H_tilde_reconstruction}, the corresponding prediction of the complete multiuser channel matrix $\hat{\mathbf{H}}_{0|t}$ can be assembled as:
\begin{equation}
    \hat{\mathbf{H}}_{0|t} = [f^{-1}(\hat{\mathbf{h}}_{1, 0|t}), f^{-1}(\hat{\mathbf{h}}_{2, 0|t}), \dots, f^{-1}(\hat{\mathbf{h}}_{K, 0|t})]. \label{eq:mu_mat_assemble}
\end{equation}

Following the signal model described in~\eqref{eq:observation_model}, we define the measurement consistency loss $\mathcal{L}_{\text{MC}}$ to penalize the discrepancy between the actual received pilot matrix $\mathbf{Y}_\mathrm{p}$ and the reconstructed pilot observations derived from our current clean prediction $\hat{\mathbf{H}}_{0|t}$:
\begin{equation}
    \mathcal{L}_{\text{MC}} = \frac{1}{2} \left\| \mathbf{Y}_\mathrm{p} - \boldsymbol{\Omega}_\mathcal{O} \hat{\mathbf{H}}_{0|t} \mathbf{P} \right\|^2_F.
\end{equation}
To steer the generative trajectories toward a solution that satisfies the actual pilot observation, we directly perform gradient descent on the predicted clean channel estimates by moving along the negative gradient of the measurement consistency loss:
\begin{equation}
    \hat{\mathbf{H}}_{0|t}(\mathbf{Y}_\mathrm{p}) = \hat{\mathbf{H}}_{0|t} - \alpha_t \nabla_{\hat{\mathbf{H}}_{0|t}} \mathcal{L}_{\text{MC}}, \label{eq:grad_descent_guidance}
\end{equation}
where $\alpha_t$ is a time-dependent step size and the gradient has a closed-form expression given by
\begin{equation}
    \nabla_{\hat{\mathbf{H}}_{0|t}} \mathcal{L}_{\text{MC}} = -\boldsymbol{\Omega}_{\mathcal{O}}^H \left( \mathbf{Y}_\mathrm{p} - \boldsymbol{\Omega}_{\mathcal{O}} \hat{\mathbf{H}}_{0|t} \mathbf{P} \right) \mathbf{P}^H. \label{eq:grad_computation}
\end{equation}
Finally, by extracting the $k$-th column of the updated matrix $\hat{\mathbf{H}}_{0|t}(\mathbf{Y}_\mathrm{p})$ (denoted by $[\cdot]_{:,k}$) and applying the forward isomorphic mapping $f$, the observation-guided clean channel estimate for each individual user at time step $t$ is given by
\begin{equation}
    \hat{\mathbf{h}}_{k, 0|t}(\mathbf{Y}_\mathrm{p}) = f\left( \left[\hat{\mathbf{H}}_{0|t}(\mathbf{Y}_\mathrm{p})\right]_{:,k} \right).
\end{equation}

\subsubsection{Practical Techniques for Robust Guidance} \label{sec:robust_guidance}

Although~\eqref{eq:grad_descent_guidance} provides a straightforward mechanism for measurement consistency guidance, a stable and robust implementation requires three practical techniques:
\begin{itemize}
    \item \textbf{Gradient Normalization:} The magnitude of $\nabla_{\hat{\mathbf{H}}_{0|t}} \mathcal{L}_{\text{MC}}$ is extremely volatile in the early generation stages (i.e., $t\to 1$), where the model's prediction is highly uncertain and incurs massive discrepancies with $\mathbf{Y}_\mathrm{p}$. We resolve this by normalizing the gradient by its Frobenius norm, so that each update magnitude is purely controlled by $\alpha_t$:
    \begin{equation}
        \hat{\mathbf{H}}_{0|t}(\mathbf{Y}_\mathrm{p}) = \hat{\mathbf{H}}_{0|t} - \alpha_t \frac{\nabla_{\hat{\mathbf{H}}_{0|t}} \mathcal{L}_{\text{MC}}}{\| \nabla_{\hat{\mathbf{H}}_{0|t}} \mathcal{L}_{\text{MC}} \|_F}. \label{eq:norm_grad_update}
    \end{equation}
    \item \textbf{Multi-Step Optimization and Interpolation:} A single gradient step per numerical integration step is typically insufficient to pull the trajectories into the desired measurement subspace. Inspired by Decomposed Diffusion Sampling (DDS)~\revb{\cite{chung2024decomposed}}, we execute $N_{\text{iter}}$ consecutive gradient steps (e.g., $N_{\text{iter}} = 3$). To mitigate over-correction while maintaining alignment with the learned channel prior, we interpolate the guided result with the unguided estimate $\hat{\mathbf{H}}_{0|t}$ to form the final combined sample:
    \begin{equation}
        \hat{\mathbf{H}}_{0|t}^{\text{guided}} = (1 - \gamma_t)\hat{\mathbf{H}}_{0|t} + \gamma_t \hat{\mathbf{H}}_{0|t}(\mathbf{Y}_\mathrm{p}).
    \end{equation}
    We set $\gamma_t = t$, so that measurement consistency dominates early while the learned prior takes over as $t \to 0$.
    \item \textbf{Injection of Stochasticity:} Since $\mathbf{Y}_\mathrm{p}$ is corrupted by AWGN, simply minimizing $\mathcal{L}_{\text{MC}}$ would cause the generation trajectories to overfit this noise. To prevent this, we inject stochasticity into the latent source noise estimate $\hat{\mathbf{z}}_{k, 1|t}$ prior to the Euler numerical integration step:
    \begin{equation}
        \hat{\mathbf{z}}_{k, 1|t}^{\text{stoch}} = \sqrt{1 - \eta_t} \hat{\mathbf{z}}_{k, 1|t} + \sqrt{\eta_t} \boldsymbol{\epsilon}, \label{eq:noise_stoch}
    \end{equation}
    where $\boldsymbol{\epsilon} \sim \mathcal{N}(0, \mathbf{I})$. The variance schedule $\eta_t = 1 - (t - \Delta t)$ increases as $t \to 0$, acting as a progressively stronger regularizer against noise overfitting. This injected stochasticity preserves the final estimation accuracy, because the Euler update~\eqref{eq:decomposed_euler} weights the noise term by $t-\Delta t$, which vanishes as $t \to 0$.
\end{itemize}

The overall algorithm of the parallel measurement consistency guided generation for multiuser channel estimation is summarized in Algorithm~\ref{alg:online_estimation}.

\begin{algorithm}[t]
\caption{Online Flow-Based Channel Estimation}
\label{alg:online_estimation}
\begin{algorithmic}[1]
\REQUIRE Pilot observation $\mathbf{Y}_\mathrm{p}$, offline-trained velocity network $\mathbf{v}_t^{\boldsymbol{\vartheta}}$, user count $K$, integration step size $\Delta t$, optimization steps $N_{\text{iter}}$, gradient step size $\alpha_t$.
\STATE \textbf{Initialize:} $\mathbf{z}_{k,1} \sim \mathcal{N}(\mathbf{0}, \mathbf{I})$ for $k = 1, \dots, K$
\FOR{$t = 1$ \textbf{down to} $0$ \textbf{with step size} $\Delta t$}
    \STATE \textit{\% 1. Flow Decomposition for Parallel Trajectories}
    \FOR{$k = 1$ \TO $K$} \label{line:flow_start}
        \STATE Compute $\hat{\mathbf{h}}_{k, 0|t} = \mathbf{z}_{k,t} - t\mathbf{v}_t^{\boldsymbol{\vartheta}}(\mathbf{z}_{k,t})$
        \STATE Compute $\hat{\mathbf{z}}_{k, 1|t} = \mathbf{z}_{k,t} + (1-t)\mathbf{v}_t^{\boldsymbol{\vartheta}}(\mathbf{z}_{k,t})$
    \ENDFOR \label{line:flow_end}
    
    \STATE \textit{\% 2. Matrix Assembly for Unified Guidance}
    \STATE Construct multiuser matrix $\hat{\mathbf{H}}_{0|t}$ via~\eqref{eq:mu_mat_assemble}
    \STATE Initialize guided estimate $\hat{\mathbf{H}}_{0|t}(\mathbf{Y}_\mathrm{p}) = \hat{\mathbf{H}}_{0|t}$
    
    \STATE \textit{\% 3. Normalized Multi-Step Guidance}
    \FOR{$i = 1$ \TO $N_{\text{iter}}$} \label{line:guide_start}
        \STATE Evaluate gradient $\nabla \mathcal{L}_{\text{MC}}$ via~\eqref{eq:grad_computation} at $\hat{\mathbf{H}}_{0|t}(\mathbf{Y}_\mathrm{p})$
        \STATE Update $\hat{\mathbf{H}}_{0|t}(\mathbf{Y}_\mathrm{p}) \gets \hat{\mathbf{H}}_{0|t}(\mathbf{Y}_\mathrm{p}) - \alpha_t \frac{\nabla \mathcal{L}_{\text{MC}}}{\| \nabla \mathcal{L}_{\text{MC}} \|_F}$
    \ENDFOR \label{line:guide_end}
    
    \STATE \textit{\% 4. Interpolation to Preserve Structural Prior}
    \STATE Compute $\hat{\mathbf{H}}_{0|t}^{\text{guided}} = (1 - \gamma_t)\hat{\mathbf{H}}_{0|t} + \gamma_t \hat{\mathbf{H}}_{0|t}(\mathbf{Y}_\mathrm{p})$
    
    \STATE \textit{\% 5. Trajectory Update via Euler Step}
    \STATE Let $t' = t - \Delta t$
    \FOR{$k = 1$ \TO $K$}
        \STATE Extract $\hat{\mathbf{h}}_{k, 0|t}^{\text{guided}} = f\Big( \big[\hat{\mathbf{H}}_{0|t}^{\text{guided}}\big]_{:,k} \Big)$
        \STATE Inject noise stochasticity via~\eqref{eq:noise_stoch}
        \STATE Compute Euler update $\mathbf{z}_{k, t'} = (1 - t')\hat{\mathbf{h}}_{k, 0|t}^{\text{guided}} + t'\hat{\mathbf{z}}_{k, 1|t}^{\text{stoch}}$
    \ENDFOR
\ENDFOR
\STATE \textbf{Output:} Estimated channels $\hat{\mathbf{h}}_k = \mathbf{z}_{k,0}, \forall k = 1, \dots, K$
\end{algorithmic}
\end{algorithm}

\begin{remark}[Bayesian Interpretation]
The measurement consistency-guided generation in Algorithm~\ref{alg:online_estimation} admits a formal Bayesian interpretation as approximate posterior sampling from $p_{\boldsymbol{\vartheta}}(\tilde{\mathbf{H}} | \mathbf{Y}_\mathrm{p}) \propto p(\mathbf{Y}_\mathrm{p} | \tilde{\mathbf{H}}) \, p_{\boldsymbol{\vartheta}}(\tilde{\mathbf{H}})$, where the joint prior factorizes as $p_{\boldsymbol{\vartheta}}(\tilde{\mathbf{H}}) = \prod_{k=1}^K p_{\boldsymbol{\vartheta}}(\mathbf{h}_k)$ under the user independence assumption established in Section~\ref{sec:sys_model}. \revb{Specifically, Algorithm~\ref{alg:online_estimation} can be viewed as a discretization of the posterior flow ODE induced by this learned prior and the observation model in~\eqref{eq:observation_model}, where the intractable gradient of the noise-perturbed likelihood is approximated by the likelihood gradient evaluated at the denoised estimates, which is proportional to $-\nabla_{\hat{\mathbf{H}}_{0|t}}\mathcal{L}_\mathrm{MC}$, and applied directly on $\hat{\mathbf{H}}_{0|t}$, a step justified under the assumption of a locally linear clean data manifold~\cite{kim2025flowdps}. The interpolation schedule $\gamma_t = t$ and the stochasticity schedule $\eta_t$ follow the choices adopted in~\cite{kim2025flowdps} for the same posterior flow, whereas the gradient normalization preserves the update direction but replaces the exact magnitude schedule with an explicitly controlled step size.} This relationship formally aligns our Stage I with the probabilistic framework introduced in Section II-C. For rigorous mathematical justification regarding this approximation, we refer interested readers to~\cite{kim2025flowdps}.
\end{remark}

\revb{In the unified two-stage framework, Algorithm~\ref{alg:online_estimation} is executed $S$ times with independent initializations and stochasticity injections, where the $s$-th run outputs a posterior multiuser channel matrix assembled as $\hat{\mathbf{H}}^{(s)} = [f^{-1}(\hat{\mathbf{h}}^{(s)}_1), \dots, f^{-1}(\hat{\mathbf{h}}^{(s)}_K)]$. Stage~I thus delivers the ensemble of posterior samples $\{\hat{\mathbf{H}}^{(s)}\}_{s=1}^{S}$ anticipated in Section~\ref{subsec:problem_formulation}, alongside the mean estimate $\hat{\mathbf{H}}_{\mathrm{mean}} = \frac{1}{S}\sum_{s=1}^{S}\hat{\mathbf{H}}^{(s)}$ obtained from the ensemble at no additional cost. The ensemble enables the computation of the empirical criterion in~\eqref{eq:empirical_utility} for the decision rule in~\eqref{eq:pool_decision}, while the mean estimate conditions the learned conditional sampler in Stage~II, as detailed in Section~\ref{sec:port_sel}, and also serves as the channel estimate whose accuracy is evaluated in Section~\ref{sec:numerical_results}.}

\subsection{Complexity Analysis} \label{sec:stage1_complexity}

We now characterize the asymptotic computational complexity of the proposed online channel estimator summarized in Algorithm~\ref{alg:online_estimation}. The overall computational cost is dominated by the U-Net forward passes and the measurement consistency gradient evaluations. Accordingly, we first quantify the complexity of a single U-Net forward pass. A standard convolutional layer operating on a spatial tensor of $N$ pixels with $D$ channels using $F \times F$ kernels requires approximately $2 N D^2 F^2$ Floating-Point Operations (FLOPs). Treating the kernel size as a constant (i.e., $F = 3$ in our implementation), the per-layer complexity simplifies to $\mathcal{O}(N D^2)$. Since our U-Net has a constant number of layers, the overall forward pass complexity is bounded by $\mathcal{O}(N D_{\max}^2)$, where $D_{\max}$ denotes the maximum channel count across all layers.

In each iteration, the algorithm performs $K$ parallel U-Net forward passes to compute the flow decomposition for each user (lines \ref{line:flow_start}--\ref{line:flow_end}), followed by $N_{\text{iter}}$ multi-step measurement consistency gradient updates on the assembled multiuser estimate (lines \ref{line:guide_start}--\ref{line:guide_end}). The gradient computation in~\eqref{eq:grad_computation} is dominated by the matrix multiplication with the orthogonal pilot codebook $\mathbf{P}$, since the selection operator $\boldsymbol{\Omega}_{\mathcal{O}}$ is binary and incurs no floating-point cost. Each gradient evaluation therefore costs $\mathcal{O}(N_{\mathcal{O}} K^2)$. Finally, the total number of iterations is given by $\mathrm{NFE} \triangleq 1/\Delta t$. Aggregating over all iterations and using the relation $N_{\mathcal{O}} = \delta N$, where $\delta$ is the sub-sampling ratio, \revb{a single run of Algorithm~\ref{alg:online_estimation} costs $O\big(\mathrm{NFE}\cdot N\cdot(KD_{\max}^2+N_{\mathrm{iter}}\delta K^2)\big)$. Since the unified framework executes Algorithm~\ref{alg:online_estimation} $S$ times to deliver the posterior ensemble, the total complexity of the proposed channel estimator is $O\big(S\cdot\mathrm{NFE}\cdot N\cdot(KD_{\max}^2+N_{\mathrm{iter}}\delta K^2)\big)$, where the $S$ runs are mutually independent and can be batched together with the $K$ per-user forward passes into a single parallel inference.}

\section{Discrete Diffusion-Based Port Selection} \label{sec:port_sel}

\revb{Building upon the channel estimates from Stage~I, Stage~II proceeds to tackle the reformulated decision problem in~\eqref{eq:map_objective}. Despite the availability of the empirical criterion~\eqref{eq:empirical_utility}, the maximization runs over the combinatorial feasible set $\mathcal{X}$, so online exhaustive search remains computationally infeasible. We therefore address this by first generating a small pool of promising port selection candidates via a learned conditional sampler, and subsequently selecting among them via the decision rule in~\eqref{eq:pool_decision}, followed by a lightweight local refinement under the same criterion.

The effectiveness of this approach heavily relies on the quality of the generated candidates. In particular, the solution obtained via~\eqref{eq:pool_decision} approaches the true global maximum of~\eqref{eq:empirical_utility} only if the pool contains near-optimal port configurations, so the sampler should concentrate its probability mass on high-utility selections under the given channel condition. To formalize this requirement, we first consider the ideal scenario of perfect channel knowledge, where the channel posterior collapses to a point mass at $\tilde{\mathbf{H}}$. The Gibbs distribution $p_\kappa(\mathbf{x}|\mathbf{Y}_\mathrm{p})$ introduced in Section~\ref{subsec:problem_formulation} then reduces to
\begin{equation}
    p_\kappa(\mathbf{x}|\tilde{\mathbf{H}}) = \frac{1}{Z} \exp\left(\kappa \cdot U(\mathbf{x}, \tilde{\mathbf{H}})\right), \label{eq:gibbs}
\end{equation}
where $Z$ is a normalizing constant. For every $\kappa > 0$, the distribution in~\eqref{eq:gibbs} inherently favors high-utility configurations, and in the limit $\kappa \to \infty$, the probability mass collapses onto the exact optimum $\mathbf{x}^*(\tilde{\mathbf{H}}) = \arg\max_{\mathbf{x} \in \mathcal{X}} U(\mathbf{x}, \tilde{\mathbf{H}})$.\footnote{The Gibbs construction serves purely as a conceptual bridge for the generative learning target. The inverse temperature scale $\kappa$ is never explicitly instantiated in the practical training pipeline.} We adopt this limiting behavior as the learning target of the Stage~II conditional sampler and employ a discrete diffusion model to learn a parametric conditional sampler $p_{\boldsymbol{\varphi}}(\mathbf{x}|\tilde{\mathbf{H}})$, whose repeated sampling passes at deployment readily generate a pool of candidates rather than a single point prediction.
}

In the remainder of this section, we first briefly introduce the fundamentals of discrete diffusion models, \revb{then elaborate on the supervised training strategy for the discrete diffusion sampler, and finally present the online decision and refinement process that completes the port selection.}

\subsection{Preliminaries on Discrete Diffusion Models} \label{sec:discrete_diffusion_overview}

Since our objective is to generate a binary mask $\mathbf{x} \in \{0,1\}^N$, the continuous flow-based diffusion framework introduced in Section~\ref{subsec:preliminaries_on_flow} is no longer applicable. Instead, we adopt a discrete diffusion modeling framework~\cite{austin2021structured}. Let $q(\mathbf{x})$ denote the target discrete distribution. Aligned with the standard denoising diffusion probability models (DDPM)~\cite{ho2020denoising}, the generative process is formulated as a $T$-step Markov chain:
\begin{equation}
    p_{\boldsymbol{\varphi}}(\mathbf{x}_{0:T}) = p(\mathbf{x}_T) \prod_{t=1}^T p_{\boldsymbol{\varphi}}(\mathbf{x}_{t-1}|\mathbf{x}_t), \label{eq:d3pm_reverse_process}
\end{equation}
where the intermediate states $\mathbf{x}_{1:T}$ share the same dimensionality as the original data $\mathbf{x}_0$. The learned distribution $p_{\boldsymbol{\varphi}}$ is then recovered by marginalizing over the discrete intermediate states: $p_{\boldsymbol{\varphi}}(\mathbf{x}_0) = \sum_{\mathbf{x}_{1:T}} p_{\boldsymbol{\varphi}}(\mathbf{x}_{0:T})$.

This generative process aims to reverse a fixed forward diffusion that progressively corrupts data into noise. By converting the binary state $\mathbf{x}$ into a one-hot categorical representation $\tilde{\mathbf{x}} = \begin{bmatrix} \mathbf{1}-\mathbf{x} & \mathbf{x} \end{bmatrix} \in \{0,1\}^{N\times 2}$ (where $\mathbf{1}$ is the all-ones vector), the forward transition is specified as
\begin{equation}
    q(\mathbf{x}_t | \mathbf{x}_{t-1}) = \text{Cat}(\mathbf{x}_t; \mathbf{p} = \tilde{\mathbf{x}}_{t-1}\mathbf{Q}_t),
\end{equation}
where $\text{Cat}(\cdot)$ denotes the categorical distribution, and $\mathbf{Q}_t \in \mathbb{R}^{2 \times 2}$ is the transition probability matrix given by
\begin{equation}
    \mathbf{Q}_t = \begin{bmatrix} 1-\beta_t & \beta_t \\ \beta_t & 1-\beta_t \end{bmatrix}.
\end{equation}
Here, the predefined noise schedule $\beta_t \in (0,1)$ is designed such that $\prod_{t=1}^T (1 - \beta_t) \approx 0$, ensuring the final state $\mathbf{x}_T$ converges to a uniform distribution. Leveraging the Markov property, the $t$-step marginal conditioned on $\mathbf{x}_0$ admits a closed form: $q(\mathbf{x}_t | \mathbf{x}_0) = \mathrm{Cat}(\mathbf{x}_t; \mathbf{p} = \tilde{\mathbf{x}}_0\bar{\mathbf{Q}}_t)$, where $\bar{\mathbf{Q}}_t = \prod_{i=1}^t \mathbf{Q}_i$ is the cumulative transition matrix.

To parameterize the reverse process in~\eqref{eq:d3pm_reverse_process}, a neural network is trained to predict the clean data distribution $p_{\boldsymbol{\varphi}}(\hat{\mathbf{x}}_0|\mathbf{x}_t)$. The reverse transition is then recovered by marginalizing over all possible predicted states:
\begin{equation}
p_{\boldsymbol{\varphi}}(\mathbf{x}_{t-1}|\mathbf{x}_t) = \sum_{\hat{\mathbf{x}}_0} q(\mathbf{x}_{t-1}|\mathbf{x}_t, \hat{\mathbf{x}}_0) p_{\boldsymbol{\varphi}}(\hat{\mathbf{x}}_0|\mathbf{x}_t). \label{eq:posterior_combining}
\end{equation}
Here, the exact posterior $q(\mathbf{x}_{t-1}|\mathbf{x}_t, \mathbf{x}_0)$ can be computed in closed form via Bayes' rule:
\begin{equation}
\begin{aligned}
    q(\mathbf{x}_{t-1}|\mathbf{x}_t, \mathbf{x}_0) &= \frac{q(\mathbf{x}_t|\mathbf{x}_{t-1}, \mathbf{x}_0)q(\mathbf{x}_{t-1}|\mathbf{x}_0)}{q(\mathbf{x}_t|\mathbf{x}_0)} \\
    &= \mathrm{Cat} \left( \mathbf{x}_{t-1}; \mathbf{p} = \frac{\tilde{\mathbf{x}}_t \mathbf{Q}_t^{\mathsf{T}} \odot \tilde{\mathbf{x}}_0 \overline{\mathbf{Q}}_{t-1}}{\tilde{\mathbf{x}}_0 \overline{\mathbf{Q}}_t \tilde{\mathbf{x}}_t^{\mathsf{T}}} \right),
\end{aligned}
\end{equation}
where $\odot$ denotes the element-wise Hadamard product.

The network parameters $\boldsymbol{\varphi}$ are optimized by minimizing the negative evidence lower bound (ELBO), which decomposes into three components:
\begin{equation}
\begin{aligned}
    \mathcal{L}_{\text{VB}}(\boldsymbol{\varphi}) &= \mathbb{E}_{q} \Bigg[ \sum_{t=2}^{T} \underbrace{D_{\text{KL}}(q(\mathbf{x}_{t-1}|\mathbf{x}_t, \mathbf{x}_0) \parallel p_{\boldsymbol{\varphi}}(\mathbf{x}_{t-1}|\mathbf{x}_t))}_{\mathcal{L}_{t-1}} \\
    &\qquad - \underbrace{\log p_{\boldsymbol{\varphi}}(\mathbf{x}_0|\mathbf{x}_1)}_{\mathcal{L}_0} + \underbrace{D_{\text{KL}}(q(\mathbf{x}_T|\mathbf{x}_0) \parallel p(\mathbf{x}_T))}_{\mathcal{L}_T} \Bigg], \label{eq:elbo}
\end{aligned}
\end{equation}
where the expectation is taken over the joint forward trajectory $q(\mathbf{x}_{0:T}) = q(\mathbf{x}_0) \prod_{t=1}^T q(\mathbf{x}_t|\mathbf{x}_{t-1})$. The prior matching term $\mathcal{L}_T$ is constant with respect to $\boldsymbol{\varphi}$ and is thus ignored. More importantly, for binary random variables, both $\mathcal{L}_0$ and $\mathcal{L}_{t-1}$ reduce to the expected cross-entropy between the predicted probability distribution $p_{\boldsymbol{\varphi}}(\hat{\mathbf{x}}_0 | \mathbf{x}_t)$ and the one-hot representation $\tilde{\mathbf{x}}_0$, which effectively serves as a deterministic ground-truth distribution. Intuitively, since~\eqref{eq:posterior_combining} constructs $p_{\boldsymbol{\varphi}}(\mathbf{x}_{t-1}|\mathbf{x}_t)$ by marginalizing the exact posterior $q(\mathbf{x}_{t-1}|\mathbf{x}_t, \mathbf{x}_0)$ over the predicted state $\hat{\mathbf{x}}_0$, matching them via $\mathcal{L}_{t-1}$ thus reduces to accurately predicting the ground truth $\tilde{\mathbf{x}}_0$. As a result, for a specific target data $\mathbf{x}_0$, the model can be tractably optimized via a surrogate binary cross-entropy (BCE) loss:
\begin{equation}
\mathcal{L}_{\text{BCE}}(\boldsymbol{\varphi}, \mathbf{x}_0) = \mathbb{E}_{t, \mathbf{x}_t} \left[ - \sum_{n=1}^N \log p_{\boldsymbol{\varphi}}(\hat{x}_{0,n} = x_{0,n} | \mathbf{x}_t) \right]. \label{eq:bce_loss}
\end{equation}
The overall objective is then to minimize $\mathbb{E}_{\mathbf{x}_0 \sim q} [\mathcal{L}_{\text{BCE}}(\boldsymbol{\varphi}, \mathbf{x}_0)]$, which can be easily optimized via SGD in model training.

\subsection{Supervised Learned Solver for Port Selection} \label{sec:sl_training}

While the general discrete diffusion framework introduced above aims to fit an unconditional target distribution $q$, \revb{Stage~II requires a conditional sampler $p_{\boldsymbol{\varphi}}(\mathbf{x}|\tilde{\mathbf{H}})$ to approximate the learning target introduced at the beginning of this section, i.e., the point mass $\delta_{\mathbf{x}^*(\tilde{\mathbf{H}})}$
obtained from~\eqref{eq:gibbs} as $\kappa\to\infty$.} Analogous to the unconditional case in~\eqref{eq:posterior_combining}, learning this conditional distribution relies on a neural network to predict the clean data distribution $p_{\boldsymbol{\varphi}}(\hat{\mathbf{x}}_0|\mathbf{x}_t, \tilde{\mathbf{H}})$ from any intermediate mask $\mathbf{x}_t$. To parameterize this conditional prediction, we utilize the same lightweight U-Net detailed in Section~\ref{sec:net_architecture}, with a slight modification in the input layer. Specifically, the $K$ real-equivalent channel vectors $\{\mathbf{h}_k\}_{k=1}^K$ and the current mask $\mathbf{x}_t$ are reshaped and concatenated into a single $(2K+1) \times N_x \times N_y$ spatial tensor for joint processing.

\revb{To establish the training objective, we follow the supervised learning (SL) paradigm of DIFUSCO~\cite{sun2023difusco} and train the conditional sampler to match this degenerate target by minimizing the expected KL divergence
\begin{equation}
    \arg \min_{\boldsymbol{\varphi}} \mathbb{E}_{\tilde{\mathbf{H}}\sim p(\tilde{\mathbf{H}})} \left[ D_{\text{KL}}\left( \delta_{\mathbf{x}^*(\tilde{\mathbf{H}})} \parallel p_{\boldsymbol{\varphi}}(\mathbf{x}|\tilde{\mathbf{H}}) \right) \right].
\end{equation}
Since the target is a point mass, this divergence reduces exactly to the negative log-likelihood of the reference optimum
\begin{equation}
    \arg\min_{\boldsymbol{\varphi}} \mathbb{E}_{\tilde{\mathbf{H}} \sim p(\tilde{\mathbf{H}})} \left[ -\log p_{\boldsymbol{\varphi}}(\mathbf{x}^*({\tilde{\mathbf{H}}}) | \tilde{\mathbf{H}}) \right], \label{eq:alternative_obj}
\end{equation}
which coincides with the log-likelihood objective that Section~\ref{sec:discrete_diffusion_overview} tractably optimizes via the negative ELBO in~\eqref{eq:elbo}. Hence, the conditional counterpart of the surrogate $\mathcal{L}_{\text{BCE}}$ derived previously in~\eqref{eq:bce_loss} can be used for training.
}

\revb{We now detail the offline construction of the labeled training pairs $(\tilde{\mathbf{H}}, \mathbf{x}^*(\tilde{\mathbf{H}}))$. To form the $i$-th pair, we assemble $\tilde{\mathbf{H}}^{(i)}$ by randomly sampling $K$ individual channels from $\mathcal{D}_{\text{train}}$ and concatenating them according to~\eqref{eq:H_tilde_reconstruction}. The reference optimum $\mathbf{x}^{*(i)} = \mathbf{x}^*(\tilde{\mathbf{H}}^{(i)})$ is then obtained via an exhaustive search over the feasible set $\mathcal{X}$, where the utility is evaluated at a fixed reference noise level of $\sigma_n^2 = 10^{-2}$, as we empirically find that the resulting optima are insensitive to this choice across the operating range. Although such a search is prohibitive online, it remains affordable in the offline phase for the moderate port count $N$ and number of RF chains $M$ considered in this work, since the utilities of all $\binom{N}{M}$ candidate subsets admit closed-form expressions and can be evaluated in parallel on a GPU.\footnote{We note that the search is independent across training pairs and thus fully parallelizable, which further extends the range of problem sizes for which exact labeling remains tractable. For substantially larger $N$ or $M$, however, exhaustive search eventually becomes infeasible even offline. Since the supervised stage only requires reference selections rather than certified optima, the exhaustive labels can then be replaced by the output of any strong search heuristic, with the learning target correspondingly relaxed from the exact optimum to the heuristic solution, while the training objective and the downstream decision and refinement process remain unchanged.} Repeating this process $N_{\text{sl}}$ times yields the supervised dataset $\mathcal{D}_{\text{sl}} = \{(\tilde{\mathbf{H}}^{(i)}, \mathbf{x}^{*(i)})\}_{i=1}^{N_{\text{sl}}}$. As a result, the entire computational burden of the exhaustive search amounts to a one-time offline cost, while the online stage involves only forward sampling passes of the trained sampler. Algorithm~\ref{alg:supervised_diffusion} summarizes the overall SL training procedure.}

\begin{algorithm}[t]
\caption{SL Training of Conditional Discrete Diffusion}
\label{alg:supervised_diffusion}
\begin{algorithmic}[1]
\REQUIRE Dataset $\mathcal{D}_{\text{sl}}$, learning rate $\zeta$, transitions $\{\bar{\mathbf{Q}}_t\}_{t=1}^T$.
\REPEAT
    \STATE Sample a pair \revb{$(\tilde{\mathbf{H}}, \mathbf{x}^*)$} from $\mathcal{D}_{\text{sl}}$
    \STATE Sample $t \sim \mathcal{U}\{1, \dots, T\}$
    \STATE Sample \revb{$\mathbf{x}_t \sim q(\mathbf{x}_t | \mathbf{x}^*) = \text{Cat}(\mathbf{x}_t; \mathbf{p} = \tilde{\mathbf{x}}^*\bar{\mathbf{Q}}_t)$}
    \STATE Compute \revb{$\mathcal{L}_{\text{BCE}} = -\sum_{n=1}^{N} \log p_{\boldsymbol{\varphi}}(\hat{x}_{0,n} = x_{n}^* | \mathbf{x}_t, \tilde{\mathbf{H}})$}
    \STATE Update network parameters: $\boldsymbol{\varphi} \leftarrow \boldsymbol{\varphi} - \zeta \nabla_{\boldsymbol{\varphi}} \mathcal{L}_{\text{BCE}}$
\UNTIL{convergence}
\end{algorithmic}
\end{algorithm}

\subsection{Online Port Selection} \label{sec:online_decision}

\revb{Once the offline training is complete, the sampler is deployed online to perform the port selection. While the sampler is trained conditioned on clean channels, at deployment it is conditioned on the mean estimate $\hat{\mathbf{H}}_{\mathrm{mean}}$ delivered by Stage~I. Any imperfection in this conditioning affects only the quality of the generated candidate pool, while the final selection is always determined by the empirical criterion~\eqref{eq:empirical_utility} through the decision and refinement steps detailed below.}

\revb{In the sequel, we drop the subscript and write $\hat{\mathbf{H}} = \hat{\mathbf{H}}_{\mathrm{mean}}$ for the conditioning input.} Since the neural network directly predicts the clean data distribution from any $\mathbf{x}_t$, we can accelerate the inference process via skipped sampling, rather than sampling the entire reverse process in~\eqref{eq:d3pm_reverse_process}. Specifically, we define a shortened sub-sequence $\{\tau_i\}_{i=0}^{T'}$ ($T' \ll T$) where $\tau_{T'} = T$ and $\tau_0 = 0$. Similar to~\eqref{eq:posterior_combining}, for any intermediate step (i.e., $i > 1$), the reverse transition from $\mathbf{x}_{\tau_i}$ to $\mathbf{x}_{\tau_{i-1}}$ follows
\begin{equation}
    p_{\boldsymbol{\varphi}}(\mathbf{x}_{\tau_{i-1}}|\mathbf{x}_{\tau_i}, \hat{\mathbf{H}}) = \sum_{\hat{\mathbf{x}}_0} q(\mathbf{x}_{\tau_{i-1}}|\mathbf{x}_{\tau_i}, \hat{\mathbf{x}}_0) p_{\boldsymbol{\varphi}}(\hat{\mathbf{x}}_0|\mathbf{x}_{\tau_i}, \hat{\mathbf{H}}). \label{eq:posterior_for_port_sel}
\end{equation}
At the final step, we apply a deterministic top-$M$ selection (denoted by $\operatorname{arg\,top}_M$) on the predicted $\hat{\mathbf{x}}_0$ probabilities instead of stochastic sampling. This guarantees that the final selected ports satisfy the FAS hardware constraint. \revb{Although the intermediate states of the independent flipping process may temporarily leave the feasible set, feasibility is required only for the final decision, which is always restored by this deterministic projection.}

\revb{A single sampling pass yields one candidate selection. To establish the candidate pool, we exploit the stochasticity of the reverse process and execute $C$ independent sampling passes in parallel, where the $c$-th pass produces the candidate $\mathbf{x}^{(c)}$ via the deterministic top-$M$ projection of its final probability prediction. The resulting pool $\{\mathbf{x}^{(c)}\}_{c=1}^{C}$ is scored by the empirical criterion, and the decision rule in~\eqref{eq:pool_decision} selects the best candidate from the pool. This selected candidate then undergoes a lightweight local refinement, in which we evaluate all single-port swaps of the current selection under $\hat{\bar{U}}_S$ and accept the best improving swap, repeating this procedure until no swap improves the criterion or a maximum of $N_r$ rounds is reached. This design follows the common practice of diffusion-based CO solvers, where multiple candidate solutions are sampled in parallel and the selected solution is further improved by a lightweight local search instead of being used directly~\cite{sun2023difusco}. Since every evaluation involved in the decision and refinement steps is a closed-form utility computation, they require no network evaluations and add negligible cost relative to the sampling passes. The complete online port selection process is summarized in Algorithm~\ref{alg:mask_generation}.}

\begin{algorithm}[t]
\caption{Online Port Selection via Discrete Diffusion}
\label{alg:mask_generation}
\begin{algorithmic}[1]
\REQUIRE \revb{Mean estimate $\hat{\mathbf{H}} = \hat{\mathbf{H}}_{\mathrm{mean}}$ and posterior samples $\{\hat{\mathbf{H}}^{(s)}\}_{s=1}^{S}$ from Stage~I}, trained model $p_{\boldsymbol{\varphi}}(\hat{\mathbf{x}}_0|\mathbf{x}_{t}, \hat{\mathbf{H}})$, sampling trajectory $\{\tau_i\}_{i=1}^{T'}$\revb{, pool size $C$, refinement rounds $N_r$}.

\FOR{\revb{$c = 1$ \TO $C$ (in parallel)}}
    \STATE Initialize $\mathbf{x}_{\tau_{T'}} \sim \text{Uniform}(\{0, 1\}^N)$
    \FOR{$i = T'$ \TO $1$}
        \STATE Compute \revb{$\mathbf{p}_0^{(c)}$}$ = p_{\boldsymbol{\varphi}}(\hat{\mathbf{x}}_0|\mathbf{x}_{\tau_{i}}, \hat{\mathbf{H}})$ via the trained model
        \IF{$i > 1$}
            \STATE Sample $\mathbf{x}_{\tau_{i-1}} \sim p_{\boldsymbol{\varphi}}(\mathbf{x}_{\tau_{i-1}}|\mathbf{x}_{\tau_{i}}, \hat{\mathbf{H}})$ given by~\eqref{eq:posterior_for_port_sel}
        \ELSE
            \STATE Construct \revb{$\mathbf{x}^{(c)}$} by setting \revb{$x^{(c)}_n$}$ = 1$ for $n \in \operatorname{arg\,top}_M (\revb{\mathbf{p}_0^{(c)}})$ and \revb{$x^{(c)}_n$}$ = 0$ otherwise
        \ENDIF
    \ENDFOR
\ENDFOR

\STATE \revb{Select the best candidate: $\hat{\mathbf{x}} \leftarrow \arg\max_{\mathbf{x} \in \{\mathbf{x}^{(c)}\}_{c=1}^{C}} \hat{\bar{U}}_S(\mathbf{x})$ according to~\eqref{eq:pool_decision}}
\FOR{\revb{$r = 1$ \TO $N_r$}}
    \STATE \revb{Evaluate $\hat{\bar{U}}_S$ on all single-port swaps of $\hat{\mathbf{x}}$}
    \IF{\revb{the best swap improves the criterion}}
        \STATE \revb{Update $\hat{\mathbf{x}}$ with the best swap}
    \ELSE
        \STATE \revb{\textbf{break}}
    \ENDIF
\ENDFOR

\RETURN Port selection strategy $\hat{\mathbf{x}}$
\end{algorithmic}
\end{algorithm}

\subsection{Complexity Analysis} \label{sec:stage2_complexity}

Owing to the shared U-Net backbone with the Stage I channel estimator, a single forward pass demands the same $\mathcal{O}(N D_{\max}^2)$ FLOPs established in Section~\ref{sec:stage1_complexity}. \revb{The candidate generation stage involves $C$ parallel skipped reverse processes with $T'$ network evaluations each, which amounts to $O(C\cdot T'\cdot N\cdot D_{\max}^2)$ FLOPs in total. The subsequent decision and refinement process requires at most $C+N_r M(N-M)$ evaluations of $\hat{\bar U}_S$, and each evaluation comprises $S$ closed-form utility computations of $O(M^3+M^2K)$ cost. The overall complexity of the proposed online port selection is therefore $O\big(C\,T'\,N\,D_{\max}^2 + S\,(C+N_r M(N-M))(M^3+M^2K)\big)$, where the first term dominates in the evaluated configuration, as discussed in Section~\ref{subsec:complexity_comparison}.}

\section{Numerical Results} \label{sec:numerical_results}

In this section, we present numerical results to evaluate the performance of the proposed unified two-stage diffusion framework for multiuser MIMO-FAS. We begin by detailing the simulation and training setups. We then validate the effectiveness of the Stage~I channel estimator in terms of accuracy and robustness, before demonstrating the system-level gains achieved by integrating both stages of the framework.

\subsection{Simulation Setup} \label{sec:simulation_setup}

We consider a narrowband system at carrier frequency $f_c = 3$~GHz. The BS FAS panel occupies a $3\lambda \times 3\lambda$ area and is uniformly quantized into $N = 25 \times 25 = 625$ ports with an ultra-dense spacing of $d=\lambda / 8$. The BS is equipped with $M=4$ active RF chains to serve $K=4$ single-FPA users. The BS FAS panel is deployed at a height of $4.0$~m, while the $K$ users at $1.5$~m are randomly distributed around the BS with radial distances uniformly drawn from $10$~m to $50$~m. Multiuser channels are generated using the QuaDRiGa simulator~\cite{jaeckel2017quadriga} under the \texttt{3GPP\_38.901\_Indoor\_NLOS} scenario. Each channel realization comprises $N_p = 20$ dominant paths, characterized by average azimuth and elevation angle spreads of $72.9^\circ$ and \revb{$39.8^\circ$}, respectively. Furthermore, essential small-scale fading parameters, including path delays, gains, and cluster distributions, are dynamically generated according to the 3GPP standard~\cite{3GPP_TR38901}.

A total of $41,000$ independent channel realizations are generated via the QuaDRiGa simulator and subsequently normalized to unit variance to facilitate network training. \revb{Since the large-scale fading varies over far coarser spatial scales than the considered FAS aperture of a few wavelengths, this normalization removes only a common scalar over all ports of each user, which leaves the port-domain channel structure unchanged and is absorbed into the operating SNR. In the subsequent evaluations, all users are accordingly assumed to experience comparable large-scale attenuations and thus a common SNR is shared.}

We then partition these realizations into a training set $\mathcal{D}_{\text{train}} = \{\mathbf{h}^{(i)}\}_{i=1}^{40,000}$ for learning the flow-based prior, and a testing set $\mathcal{D}_{\text{test}}$ of $1,000$ samples for downstream evaluation. \revb{To train the discrete diffusion-based port selector, we further construct a paired dataset $\mathcal{D}_{\text{sl}} = \{(\tilde{\mathbf{H}}^{(i)}, \mathbf{x}^{*(i)})\}_{i=1}^{N_{\text{sl}}}$ with $N_{\text{sl}} = 10{,}000$ exhaustive-search labels according to the procedure outlined in Section~\ref{sec:sl_training}.} The discrete diffusion process utilizes \revb{$T = 1000$} steps with a cosine noise schedule~\cite{nichol2021improved}. \revb{The flow-based prior is trained for $250$ epochs and the conditional port selection sampler for $100$ epochs, both} using the Adam optimizer with a learning rate of $\zeta = 10^{-4}$.

\subsection{Impact of Sampling Steps on Channel Estimation Accuracy} \label{sec:sampling_steps}

\begin{figure}[t]
    \centering
    \includegraphics[width=0.8\linewidth]{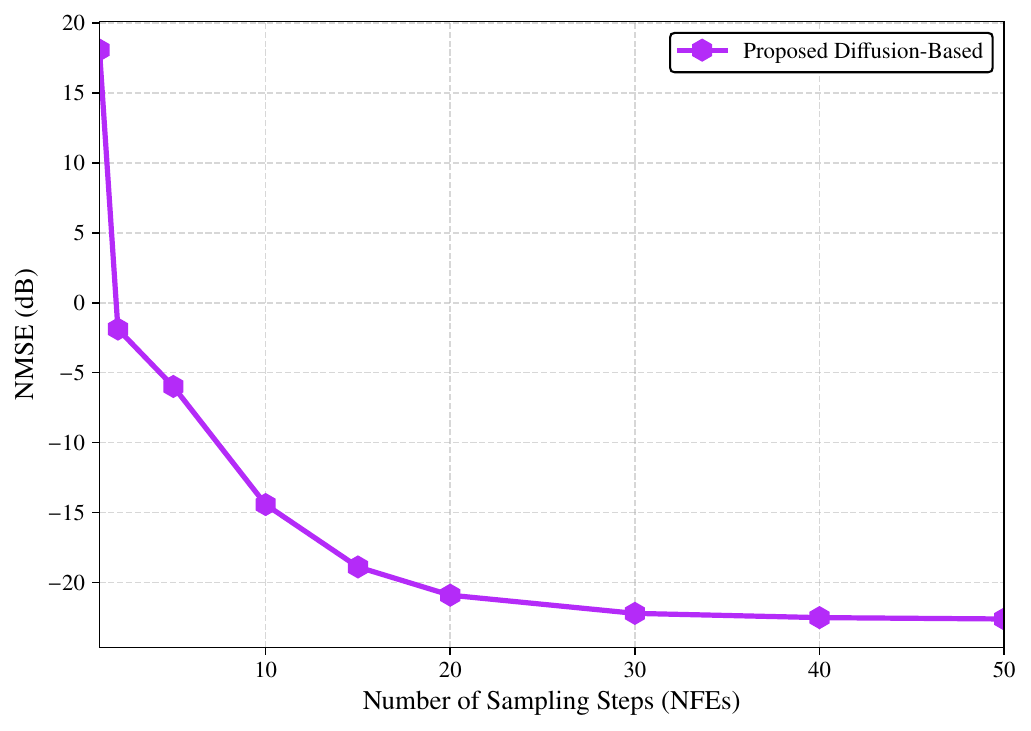}
    \caption{NMSE of the proposed channel estimator under different NFEs at $N_{\mathcal{O}} = 121$ and $20$~dB SNR.}
    \label{fig:nmse_vs_nfe}
\end{figure}

For the Stage~I channel estimation evaluations, unless otherwise specified, we fix the number of observed ports to $N_{\mathcal{O}} = 121$ and the signal-to-noise ratio (SNR) at $20$~dB. \revb{Each reported point is averaged over $100$ Monte Carlo runs.} The orthogonal pilot codebook $\mathbf{P}$ is deterministically constructed from a Hadamard matrix. The estimation accuracy is measured by the normalized mean square error (NMSE), defined as
\begin{equation}
    \text{NMSE} = \mathbb{E} \left[ \frac{\|\hat{\mathbf{H}} - \tilde{\mathbf{H}}\|_F^2}{\|\tilde{\mathbf{H}}\|_F^2} \right],
\end{equation}
where $\tilde{\mathbf{H}}$ and $\hat{\mathbf{H}}$ represent the ground-truth and estimated multiuser channel matrices, respectively. Given that the generated channel samples are normalized, the SNR simplifies to $\text{SNR} = 1/\sigma_n^2$. Furthermore, for the proposed flow-based channel estimator detailed in Algorithm~\ref{alg:online_estimation}, the number of guidance iterations is set to $N_{\text{iter}} = 3$, \revb{and the gradient step size $\alpha_t$ is tuned once for each evaluated operating point and then held fixed. Unless otherwise specified, each reported channel estimate is formed as the average of $S = 4$ posterior samples, consistent with the empirical decision criterion in Section~\ref{subsec:problem_formulation}.}

Fig.~\ref{fig:nmse_vs_nfe} illustrates the NMSE performance of the proposed flow-based channel estimator against the NFEs. The results demonstrate a sharp reduction in NMSE as the NFE increases from a highly coarse sampling grid, before quickly stabilizing beyond an NFE of approximately $20$. This indicates that a denser sampling trajectory (i.e., a larger NFE) yields marginal estimation gains at the expense of a proportional increase in inference latency. This rapid saturation demonstrates the promising potential of our flow-based generative approach for real-time, high-dimensional multiuser FAS channel estimation. \revb{Accordingly, the subsequent evaluations adopt an NFE of 50 by default, which lies comfortably within the saturation region.}

\subsection{Robustness of Channel Estimation Across SNR Regimes} \label{sec:snr_robustness}

For fair comparisons with existing FAS channel estimation techniques, which are primarily designed for SISO-FAS scenarios, we first decouple the multiuser channel estimation problem into $K$ individual SISO recovery tasks. This is achieved by right-multiplying the observations in~\eqref{eq:observation_model} by the pilot pseudo-inverse, i.e., $\mathbf{Y}_\mathrm{p}\mathbf{P}^\dagger$. If full array observation is available ($N_\mathcal{O} = N$), this least squares (LS) estimator can directly serve as a fundamental performance benchmark. For practical sub-sampled recovery scenarios ($N_\mathcal{O} < N$), we evaluate three baselines:
\begin{itemize}
    \item \textbf{OMP}: The classical Orthogonal Matching Pursuit (OMP) algorithm exploits the inherent channel sparsity of $\tilde{\mathbf{h}}_k$ in the angular domain, where we construct an over-complete dictionary with a resolution of $G = 50$ grids for both azimuth and elevation AoAs.
    \item \textbf{SBL}: The Sparse Bayesian Learning (SBL) method proposed in~\cite{xu2024sparse} utilizes the expectation-maximization (EM) algorithm to learn a parameterized Gaussian prior over the sparse angular representation of $\tilde{\mathbf{h}}_k$ based on the same dictionary as OMP.
    \item \textbf{LMMSE}: The Linear Minimum Mean Square Error (LMMSE) estimator assumes a simplified Gaussian prior $\tilde{\mathbf{h}}_k \sim \mathcal{CN}(\mathbf{0}, \mathbf{R}_{\mathbf{h}})$, where the covariance matrix $\mathbf{R}_{\mathbf{h}}$ is empirically computed from the $40,000$ training samples.
\end{itemize}

\begin{figure}[t]
    \centering
    \includegraphics[width=0.8\linewidth]{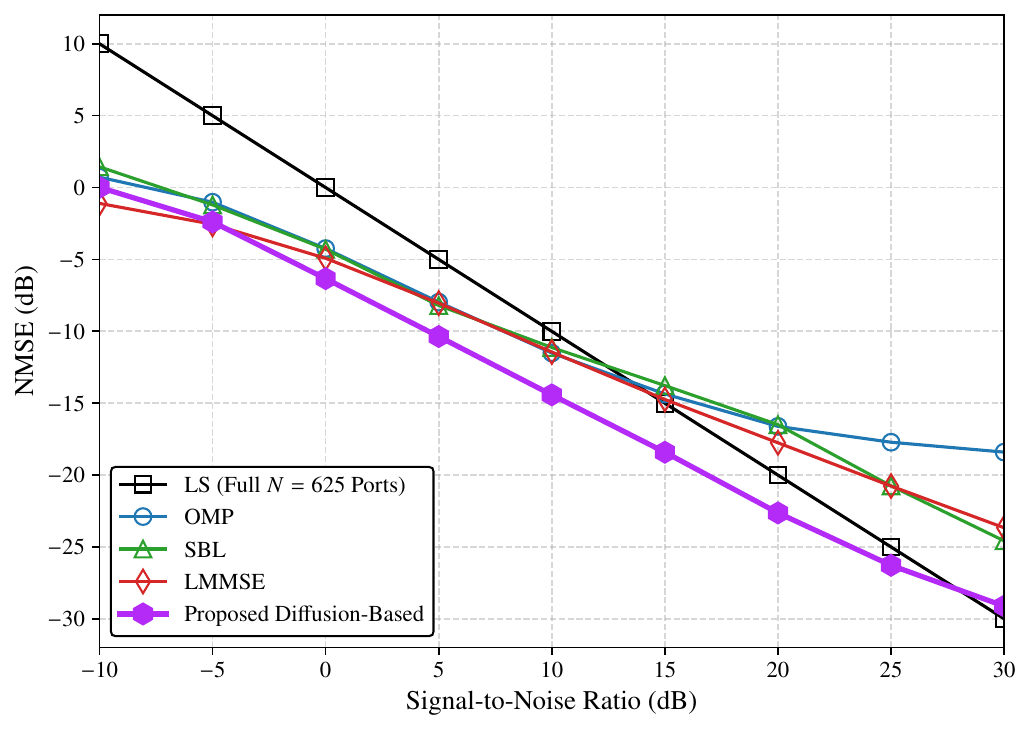}
    \caption{NMSE performance versus SNR, where all sub-sampled schemes operate at $N_{\mathcal{O}} = 121$.}
    \label{fig:nmse_vs_snr}
\end{figure}

Fig.~\ref{fig:nmse_vs_snr} illustrates the NMSE comparison for SNRs from $-10$ to $30$~dB\revb{, where the proposed estimator and all sub-sampled baselines operate at $N_{\mathcal{O}} = 121$, while the LS reference requires observing the full array}. Although the full-observation LS estimator achieves competitive accuracy at high SNRs, it requires a prohibitive switching overhead across all $N$ ports. In contrast, the proposed flow-based estimator significantly outperforms all sub-sampled baselines \revb{from $0$ dB onward, while the LMMSE baseline still dominates in the severely noise-limited regime}. These performance gaps further widen under more restrictive observations, as elaborated in the following subsection.

\subsection{Impact of Sub-sampling Ratio} \label{sec:subsampling_ratio}

\begin{figure}[t]
    \centering
    \includegraphics[width=0.8\linewidth]{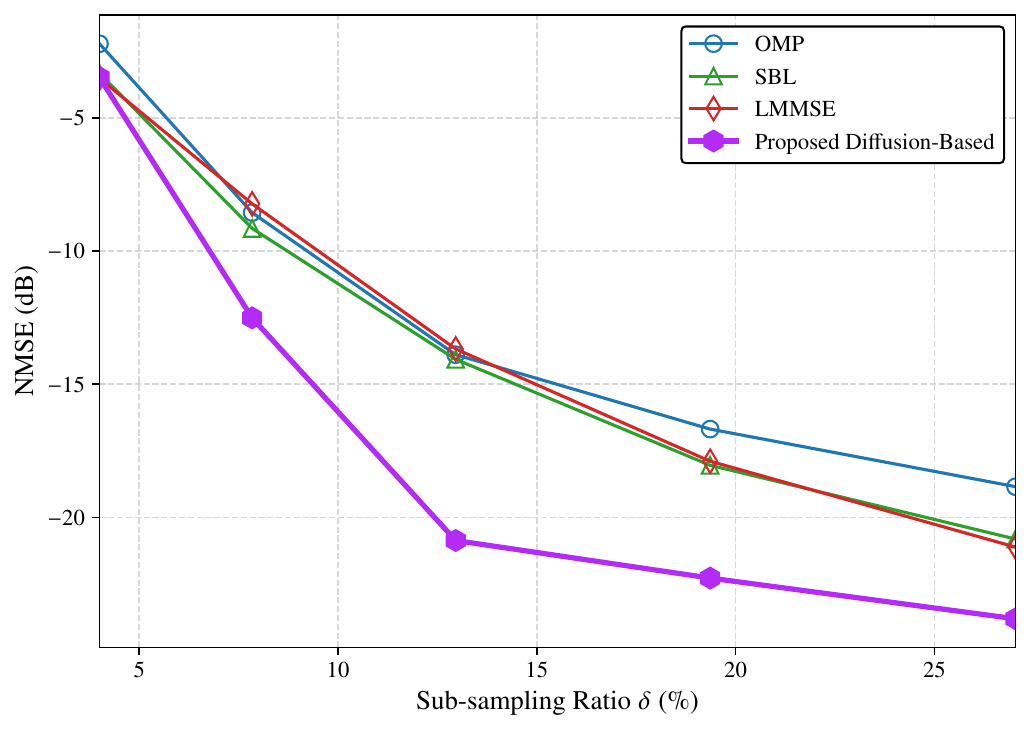}
    \caption{NMSE performance versus sub-sampling ratio $\delta$ under $20$~dB SNR.}
    \label{fig:nmse_vs_subsampling}
\end{figure}

To further investigate the channel estimation performance under severely observation-limited conditions, we evaluate the NMSE against varying sub-sampling ratios $\delta = N_{\mathcal{O}}/N$ at a fixed $\text{SNR} = 20$~dB. The LS estimator is excluded from this evaluation since it requires full array observation, i.e., $\delta = 100\%$. Fig.~\ref{fig:nmse_vs_subsampling} reveals a substantial performance gap in the severely under-sampled regime. Specifically, as $\delta$ increases from $4\%$ to \revb{$13\%$}, our flow-based estimator exhibits a rapid NMSE drop, significantly outperforming all other baselines.

We attribute this remarkable performance gain to the fact that our flow-based channel estimator implicitly learns the intricate true prior distribution of the FAS channels, enabling robust reconstruction even from severely limited spatial observations. In contrast, CS-based schemes (OMP and SBL) suffer from highly ill-conditioned measurement matrices at low sub-sampling ratios, preventing them from accurately resolving the sparse angular paths. In addition, the LMMSE approach relies on an over-simplified Gaussian prior that is insufficient for faithful recovery under limited observations.

\revb{This sharp transition can be interpreted from the perspective of spatial degrees of freedom (DoF). The $N=625$ dimensional channel of each individual user is a heavily oversampled representation of a bandlimited electromagnetic field over the $3\lambda\times3\lambda$ aperture, which possesses at most $\lceil\pi W_xW_y\rceil=29$ significant spatial DoF under isotropic scattering according to the Fourier plane-wave expansion~\cite{pizzo2022fourier}. The two smallest observation budgets in Fig.~\ref{fig:nmse_vs_subsampling}, i.e., $N_\mathcal{O}=25$ and $N_\mathcal{O}=49$, lie on the two sides of this threshold, which explains why the estimation accuracy collapses at $\delta=4\%$ but then recovers rapidly once the number of observed ports exceeds the effective channel dimensionality. The placement of the observed ports is chosen accordingly. For $N_\mathcal{O}\ge 49$, the observed ports are placed on a uniform grid whose spacing of at most $0.5\lambda$ satisfies the spatial Nyquist criterion, and we find this regular placement preferable in our experiments. At the smallest budget $N_\mathcal{O}=25$, a uniform grid would lead to a spacing of $0.75\lambda$ and thus introduce spatial aliasing, so a random pattern is adopted instead.}

\begin{figure}[t]
    \centering
    \includegraphics[width=0.8\linewidth]{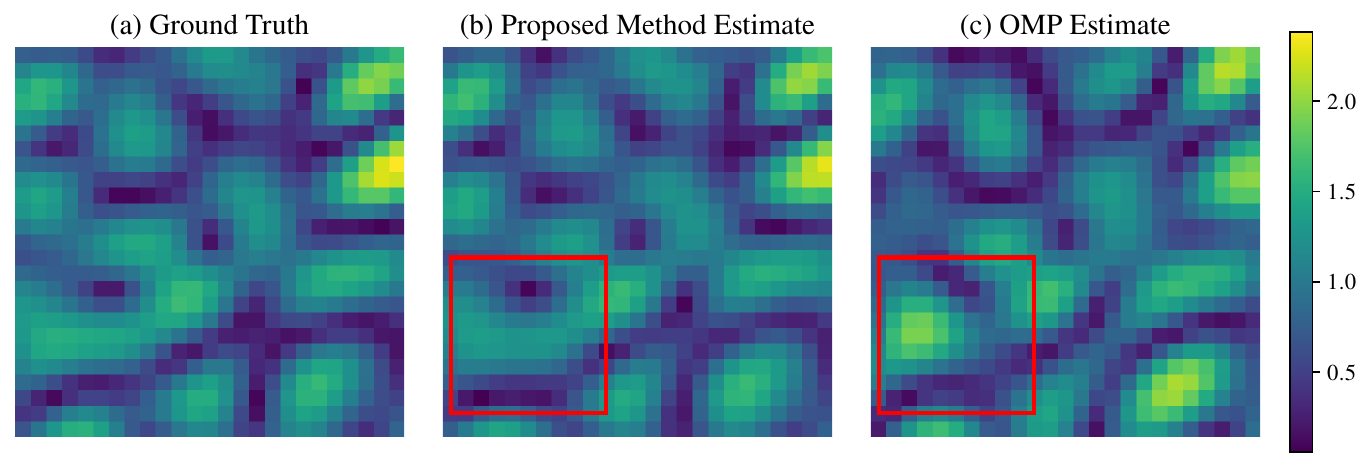}
    \caption{Visual comparison of estimated channel magnitudes ($\text{SNR} = 20$~dB, $\delta \approx 8\%$). Red boxes highlight structural estimation errors in the OMP baseline, which are effectively mitigated by the proposed method.}
    \label{fig:comparison}
\end{figure}

To provide deeper insights into the reconstruction quality, Fig.~\ref{fig:comparison} presents a visual comparison of the estimated channels for a representative user at $\text{SNR} = 20$~dB and $\delta \approx 8\%$. As highlighted by the red boxes, the OMP estimate exhibits significant structural estimation errors, while the proposed channel estimator faithfully recovers the true spatial variations. Beyond numerical accuracy, preserving this spatial pattern integrity is also critical, since any structural artifact will inevitably mislead the downstream port selection, ultimately resulting in a severe degradation of the achievable capacity.

\subsection{Robustness Under Distribution Mismatch} \label{sec:mismatch_robustness}

\begin{figure}[t]
    \centering
    \includegraphics[width=0.8\linewidth]{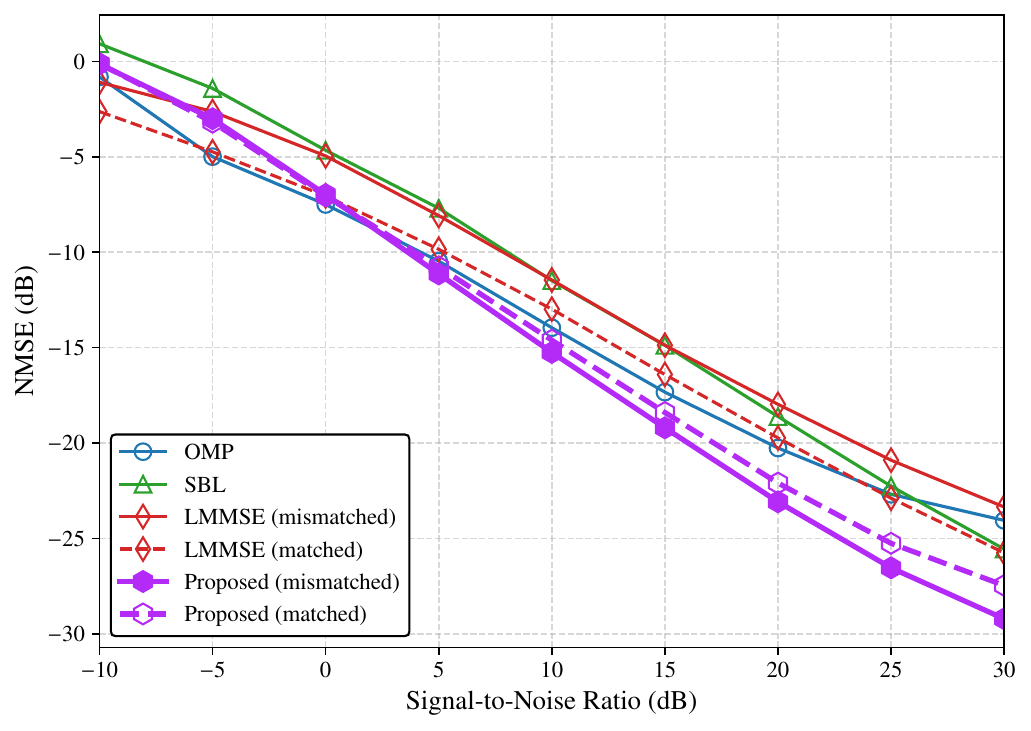}
    \caption{NMSE versus SNR at $N_{\mathcal{O}} = 121$ under scenario mismatch.}
    \label{fig:robustness_mismatch}
\end{figure}

\revb{
Since the flow-based prior is learned from a specific channel distribution, we further examine its robustness when the deployment distribution deviates from the training distribution. To this end, we generate a new test set using the LOS counterpart of the \texttt{3GPP\_38.901\_Indoor\_NLOS} scenario adopted for training, while keeping all other generation parameters unchanged, and evaluate the trained estimator on it without any retraining or fine-tuning. Fig.~\ref{fig:robustness_mismatch} plots the NMSE against the SNR at $N_{\mathcal{O}} = 121$. As a reference, we retrain the estimator on an equally sized set of LOS channel samples using the identical architecture, training budget, and hyperparameters, which yields the matched counterpart of the deployed mismatched estimator, so that the only difference between the two lies in the training distribution.

As shown in Fig.~\ref{fig:robustness_mismatch}, the mismatched estimator leads all evaluated baselines from moderate SNR onward. At very low SNRs, the strongest conventional baselines still prevail, as expected in the noise-dominated regime where the pilot observations provide limited information. More notably, the mismatched estimator performs comparably to its retrained counterpart at low SNRs and even exceeds it at moderate-to-high SNRs, so retraining on the deployment scenario brings no benefit in our setting.

The two LMMSE curves in the same figure verify the validity of the LOS dataset and the evaluation pipeline, as the matched covariance outperforms the mismatched one over the entire evaluated SNR range, which is consistent with classical linear estimation theory. The retrained estimator also remains competitive, as it still surpasses every baseline at moderate and high SNRs. We therefore attribute the advantage of the mismatched estimator to the diversity of the training data. The NLOS training set exposes the network to a much richer family of multipath structures and thereby provides a stronger learning signal for the spatial correlation patterns shared by both scenarios, whereas LOS channels dominated by a single strong path offer little such structural variability.
}

\subsection{Minimum Achievable Rate With Estimated Channel} \label{sec:achievable_rate}

\begin{figure}[t]
    \centering
    \includegraphics[width=0.8\linewidth]{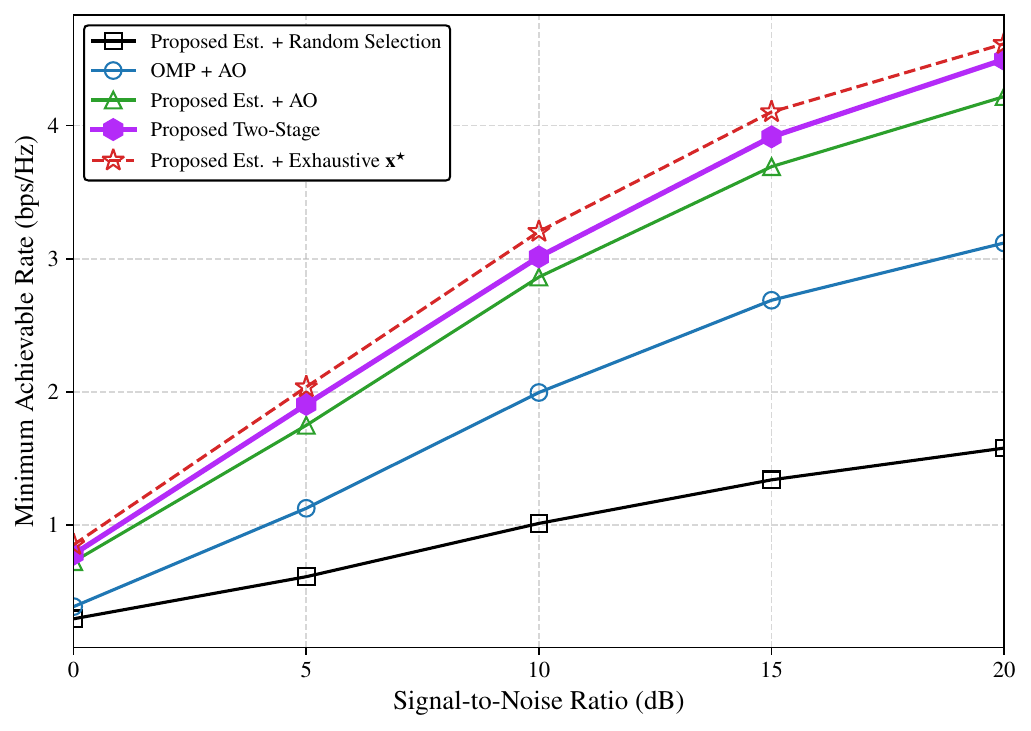}
    \caption{Minimum Achievable Rate versus SNR.}
    \label{fig:min_achievable_rate}
\end{figure}

Finally, Fig.~\ref{fig:min_achievable_rate} evaluates the overall system-level performance by plotting the deployed minimum achievable rate versus SNR. \revb{Each scheme performs port selection based on its own channel estimate, and the combining matrix is obtained by applying \eqref{eq:mmse_receiver} to the same estimate restricted to the selected ports. The resulting minimum rate is then evaluated on the true channel at the same operating SNR. Specifically, the SINR in \eqref{eq:sinr} is computed with the estimate-based combiner in place of $\mathbf{w}_k(\mathbf{x},\tilde{\mathbf{H}})$, while the effective channels $\bar{\mathbf{h}}_k$ remain those of the true channel. This end-to-end evaluation exposes how estimation errors propagate into both the port selection and the final combining, and it assesses the proposed two-stage framework, which is derived as a tractable approximation of \eqref{eq:map_objective}, against conventional cascaded designs under identical deployment conditions. Five schemes are compared as follows.}

\begin{itemize}
    \item \revb{\textbf{Proposed Est. + Random Selection}: $M=4$ active ports drawn uniformly at random, deployed with the combiner computed from the proposed Stage~I estimate, which serves as a selection-free floor.}
    \item \revb{\textbf{OMP + AO}: Channel estimation via OMP, followed by the port-optimization block of the AO solver~\cite{cheng2024exploiting}, which performs element-wise discrete port updates under the closed-form MMSE combiner.}
    \item \revb{\textbf{Proposed Est. + AO}: The same AO port selection operating on the proposed Stage~I estimate.}
    \item \revb{\textbf{Proposed Two-Stage}: Our complete framework, where the discrete diffusion sampler proposes $C=16$ candidates via $T'=100$ skipped reverse steps, and the final decision maximizes the empirical criterion in~\eqref{eq:empirical_utility} ($S=4$) over the pool, followed by at most $N_r=5$ rounds of local refinement.}
    \item \revb{\textbf{Proposed Est. + Exhaustive $\mathbf{x}^\star$}: The port selection that exhaustively maximizes the point-estimate utility $U(\mathbf{x},\hat{\mathbf{H}}_{\mathrm{mean}})$ over all $\binom{N}{M}$ subsets, deployed with the same combiner.\footnote{This reference is computationally infeasible online, since each channel instance requires $\binom{625}{4}\approx 6.3\times 10^{9}$ closed-form utility evaluations and consumes about ten seconds even on a data-center GPU. It represents the computational limit of the conventional cascaded pipeline, since no scheme that optimizes the port selection against a single estimated channel can attain a higher utility on that same estimate.}}
\end{itemize}

\revb{As shown in Fig.~\ref{fig:min_achievable_rate}, the proposed two-stage framework achieves the highest deployed rate among all practical schemes at every evaluated SNR. Pairwise comparisons among these schemes further isolate the contribution of each stage. The comparison between \textbf{OMP + AO} and \textbf{Proposed Est. + AO} isolates the estimation stage. When the same AO selector operates on the proposed Stage~I estimate instead, the rate improves substantially at all SNRs, as structural estimation errors in the OMP estimate directly mislead the downstream selection. The comparison between \textbf{Proposed Est. + AO} and \textbf{Proposed Two-Stage} isolates the selection stage on the identical Stage~I output. The advantage grows with the SNR, because a more accurate estimate allows the decision rule in \eqref{eq:pool_decision} to exploit the candidate pool more reliably, whereas the element-wise AO updates remain trapped in locally optimal selections. This loss is inherent to the local search itself and thus cannot be alleviated by a more accurate estimate.}

\revb{Among the five schemes, the exhaustive reference marks the performance limit of the conventional cascaded pipeline, since it exactly solves the max-min selection problem in~\eqref{eq:port_sel_orig} with the true channel replaced by the posterior mean estimate $\hat{\mathbf{H}}_{\mathrm{mean}}$ from Stage~I. As shown in Fig.~\ref{fig:min_achievable_rate}, the proposed two-stage framework is the closest practical scheme to this reference and achieves more than $90\%$ of its deployed rate across the entire evaluated SNR range. These results confirm that the proposed framework approaches this limit at a feasible online cost.}

\subsection{Complexity Comparison} \label{subsec:complexity_comparison}

Tables \ref{tab:complexity_stage1} and \ref{tab:complexity_stage2} compare the per-stage computational complexity of the proposed framework with baselines, where $I_{\mathrm{OMP}}$, $I_{\mathrm{SBL}}$, and $I_{\mathrm{AO}}$ denote the iteration counts of OMP, SBL, and AO, respectively. Notably, the proposed Stage I estimator scales linearly with $N$, matching OMP while bypassing the prohibitive $\mathcal{O}(N^3)$ matrix inversion bottleneck of SBL and LMMSE. \revb{This efficient scaling also extends to Stage II, where both terms of the overall complexity grow linearly with $N$. The decision and refinement process shares the single-port swap structure of the AO baseline and is capped at $N_r$ rounds, so its closed-form utility computations grow only linearly with $N$ and remain negligible against the combinatorial $\binom{N}{M}$ evaluations of an online exhaustive search. The overall online cost is instead dominated by the $C\cdot T'$ network forward passes of the candidate generation stage. Consequently, the unified two-stage framework remains compatible with large-scale multiuser MIMO-FAS deployments.}

\begin{table}[!t]
    \renewcommand{\arraystretch}{1.2}
    \footnotesize
    \caption{Stage~I Computational Complexity Comparison}
    \label{tab:complexity_stage1}
    \centering
    \begin{tabular}{ll}
        \hline\hline
        \textbf{Algorithm} & \textbf{Computational Complexity} \\
        \hline
        OMP & $\mathcal{O}(K \cdot I_{\mathrm{OMP}} \cdot \delta N G^2)$ \\
        SBL~\cite{xu2024sparse} & $\mathcal{O}(K \cdot I_{\mathrm{SBL}} \cdot (\delta^2 N^2 G^2 + \delta^3 N^3))$ \\
        LMMSE & $\mathcal{O}(K \cdot (\delta^3 N^3 + \delta N^2))$ \\
        \textbf{Proposed} & \revb{$O\big(S\cdot\mathrm{NFE}\cdot N\cdot(KD_{\max}^2+N_{\mathrm{iter}}\delta K^2)\big)$} \\
        \hline\hline
    \end{tabular}
\end{table}

\begin{table}[!t]
    \renewcommand{\arraystretch}{1.2}
    \footnotesize
    \caption{Stage~II Computational Complexity Comparison}
    \label{tab:complexity_stage2}
    \centering
    \begin{tabular}{ll}
        \hline\hline
        \textbf{Algorithm} & \textbf{Computational Complexity} \\
        \hline
        AO~\cite{cheng2024exploiting} & $O\big(I_{\mathrm{AO}}\cdot M(N-M)\cdot(M^3+M^2K)\big)$ \\
        \textbf{Proposed} & \revb{$O\big(C\,T'\,N\,D_{\max}^2 + S\,N_r\,M(N-M)\cdot(M^3+M^2K)\big)$} \\
        \hline\hline
    \end{tabular}
\end{table}

\vspace{-0.1cm}

\section{Conclusion} \label{sec:conclusion}

In this paper, we proposed a unified two-stage diffusion framework to solve the fundamental challenges of channel estimation and port selection in multiuser MIMO-FAS. By casting the joint task as a single MAP inference problem \revb{equivalent to maximizing the posterior expected utility of the port selection, and approximating this objective through two sequential stages}, the proposed framework establishes a principled probabilistic bridge between \revb{estimation and selection}, realized by a flow-based channel posterior sampler and a discrete diffusion port selector \revb{whose final decision maximizes the empirical posterior expected utility over the generated candidates}. Numerical results confirm that our framework \revb{attains over $90\%$ of the deployed rate of the computationally expensive exhaustive search over the estimated channel at a feasible online cost, and substantially improves} the minimum achievable rate over \revb{practical cascaded} baselines, driven by its exceptional robustness against severely sub-sampled pilots and realistic estimation errors. In summary, this work establishes generative diffusion models as a compelling solution for tackling high-dimensional, non-convex physical-layer challenges, offering a practical and scalable methodology for large-scale MIMO-FAS deployments.

\revb{Finally, we comment on the scope of the learned channel prior. The prior is specific to the channel distribution on which it is trained in two respects. First, it is matched to the port spacing of the training configuration, so deployment at a different spacing, particularly a coarser one, would require retraining. Extending the framework toward resolution-invariant channel estimation is part of our ongoing work. Second, the training channels characterize propagation only and exclude hardware effects such as mutual coupling among densely spaced ports. Since such effects act as a linear transformation of the channel at the antenna ports, they can in principle be absorbed by the prior once it is trained on electromagnetically simulated or measured channels. Accordingly, retraining and validating the framework under such conditions are left for future work.}


\bibliographystyle{IEEEtran}
\bibliography{main}

\end{document}